\pdfoutput=1

\documentclass[11pt,twoside,a4paper,cmspaper,final,collab]{cms-tdr}
\def\svnVersion{428788P}\def\cmsCernNoTag{CERN-EP-2017-109}\def\cmsCernDate{\today}\def\cmsMessage{Published in the Journal of High Energy Physics as \href{http://dx.doi.org/10.1007/JHEP10(2017)019}{\doi{10.1007/JHEP10(2017)019.}}}

\begin{document}\cmsNoteHeader{SUS-16-051}

\hyphenation{had-ron-i-za-tion}
\hyphenation{cal-or-i-me-ter}
\hyphenation{de-vices}
\RCS$Revision: 428720 $
\RCS$HeadURL: svn+ssh://svn.cern.ch/reps/tdr2/papers/SUS-16-051/trunk/SUS-16-051.tex $
\RCS$Id: SUS-16-051.tex 428720 2017-10-10 14:32:33Z echabert $

\newcommand{\wjets}{\ensuremath{\PW\text{+jets}}\xspace}
\newcommand{\zjets}{\ensuremath{\PZ\text{+jets}}\xspace}
\newcommand{\gjets}{\ensuremath{\Pgg\text{+jets}}\xspace}
\newcommand{\MT}{\ensuremath{M_{\mathrm{T}}}\xspace}
\newcommand{\HTmiss}{\ensuremath{H_{\mathrm{T}}^{\text{miss}}}\xspace}
\newcommand{\tmod}{\ensuremath{t_{\mathrm{mod}}}\xspace}
\newcommand{\Mlb}{\ensuremath{M_{\ell \PQb}}\xspace}
\newcommand{\minDPhiMETjet}{\ensuremath{\min\Delta\phi(J_{1,2},\MET)}\xspace}
\newcommand{\NJ}{\ensuremath{N_{\mathrm{J}}}\xspace}
\newcommand{\pp}{\ensuremath{\Pp\Pp}\xspace}
\newcommand{\Lint}{35.9\fbinv\xspace}

\cmsNoteHeader{SUS-16-051} \title{Search for  top squark pair production in pp collisions at
$\sqrt{s}=13\TeV$ using single lepton events}

\date{\today}

\abstract{
A search for top squark pair production in pp collisions at $\sqrt{s}=13\TeV$ is performed using events with a single isolated electron or muon, jets, and a large transverse momentum imbalance.
The results are based on data collected in 2016 with the CMS detector at the LHC, corresponding to an integrated luminosity of 35.9\fbinv.
No significant excess of events is observed above the expectation from standard model processes.
Exclusion limits are set in the context of supersymmetric models of
pair production of top squarks that decay either to a top quark and a
neutralino or to a bottom quark and a chargino.  Depending on the
details of the model, we exclude top squarks with masses as high as
1120\GeV. Detailed information is also provided to facilitate theoretical interpretations in other scenarios of physics beyond the standard model.
}

\hypersetup{pdfauthor={CMS Collaboration},pdftitle={Search for top squark pair production in pp collisions at
sqrt(s)=13 TeV using single lepton events},pdfsubject={CMS},pdfkeywords={CMS, physics, SUSY, stop, squark}}

\maketitle
\section{Introduction}
\label{sec:intro}

\begin{figure}[!htpb]
\centering
\includegraphics[width=0.45\textwidth]{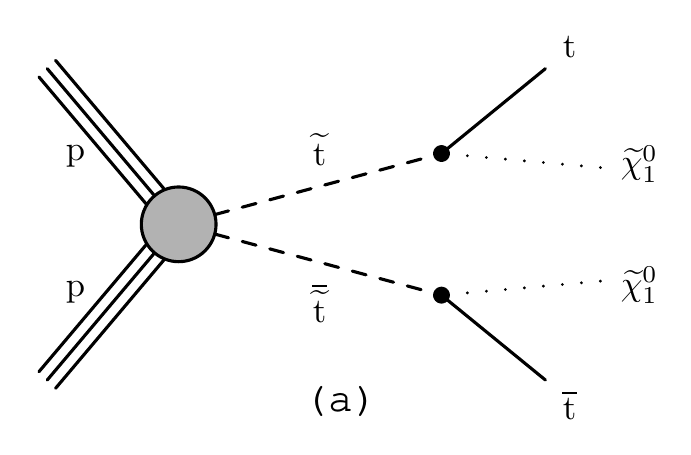}\hskip1cm
\includegraphics[width=0.45\textwidth]{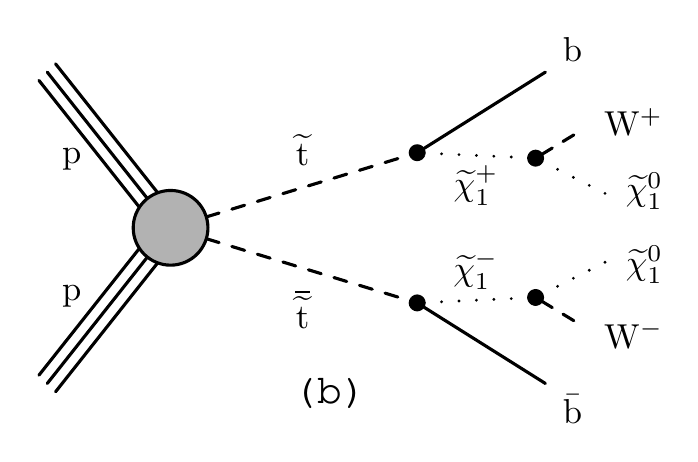}\vskip 0.5cm
\includegraphics[width=0.45\textwidth]{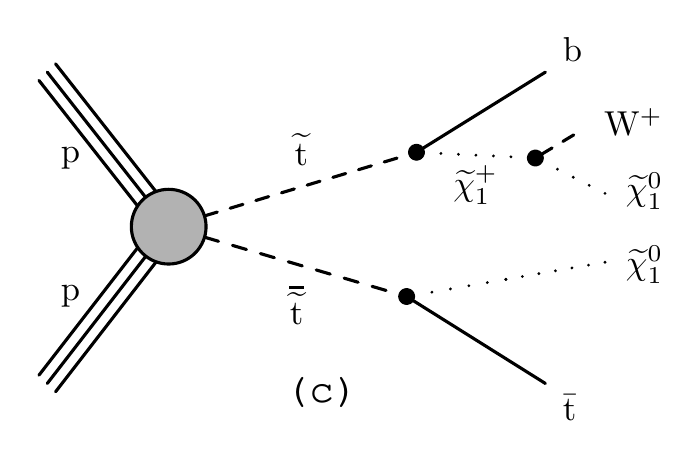}\caption{\label{fig:diagram}Simplified-models diagrams corresponding to top squark pair
  production, followed by the specific decay modes targeted in this
  paper. (a) $\Pp \Pp \to \PSQt\,\PASQt \to \PQt \PSGczDo ~ \PAQt \PSGczDo$; (b) $\Pp \Pp \to \PSQt\,\PASQt \to \PQb \PSGcpDo \PAQb \PSGcmDo$; (c) $\Pp \Pp \to \PSQt\,\PASQt \to \PQb \PSGcpDo \PAQt \PSGczDo$. Charge-conjugate decays are implied.}

\end{figure}

Supersymmetry (SUSY) ~\cite{Ramond:1971gb,Golfand:1971iw,Neveu:1971rx,Volkov:1972jx,Wess:1973kz,Wess:1974tw,Fayet:1974pd,Nilles:1983ge}
is an extension of the standard model (SM) that postulates the
existence of a superpartner for every SM particle with the same gauge quantum numbers but differing by one half-unit of spin.
The search for a low mass top squark, the scalar partner of the top
quark, is of particular interest following the discovery of a Higgs
boson \cite{Aad:2012tfa,Chatrchyan:2012xdj,Chatrchyan:2012tx}, as it would
substantially contribute to the cancellation of the divergent loop
corrections to the Higgs boson mass,
providing a possible solution to the hierarchy problem \cite{Papucci:2011wy,Barbieri:2009ev,Dimopoulos:1995mi}.
We present results of a search for top squark pair production in the
final state with a single lepton ($\ell=  \Pe$ or \Pgm) with high transverse momentum (\pt),
jets, and significant \pt imbalance.  Dedicated top squark
searches have been carried out
by the ATLAS~\cite{ATLASstop1L2015} and CMS~\cite{Sirunyan:2016jpr,Khachatryan:2017rhw} collaborations based on 13\TeV proton-proton~(pp) collisions at the CERN LHC, with data sets corresponding to integrated
luminosities of 3.2 and 2.3\fbinv, respectively.
In this paper we report on an extension of the search of Ref.~\cite{Sirunyan:2016jpr} in the single-lepton final state that exploits
the data sample collected with the CMS detector~\cite{JINST} in 2016, corresponding to the much larger integrated luminosity of \Lint.
We find no evidence for an excess of events above the expected background from standard model processes, and interpret the results as limits on simplified models~\cite{ArkaniHamed:2007fw,Alwall:2008ag,Alwall:2008va,Alves:2011wf} of the pair production of top squarks (\PSQt) decaying
into top quarks and neutralinos (\PSGczDo) and/or bottom quarks and
charginos ($\ensuremath{\widetilde{\chi}^{\pm}_{1}}\xspace$), as shown in Fig.~\ref{fig:diagram}.  We take the \PSGczDo to be the lightest supersymmetric particle (LSP) and to
be stable.

\section{The CMS detector}

The central feature of the CMS apparatus is a superconducting solenoid of 6\unit{m} internal diameter,
providing a magnetic field of 3.8\unit{T}. Within the solenoid volume are a silicon pixel and strip tracker, a
lead tungstate crystal electromagnetic calorimeter, and a brass and scintillator hadron calorimeter, each composed of
a barrel and two endcap sections. Forward calorimeters extend the pseudorapidity ($\eta$) coverage provided by the
 barrel and endcap detectors. Muons are measured in gas-ionization detectors embedded in the steel flux-return yoke outside
the solenoid. The first level of the CMS trigger system, composed of custom hardware processors, uses information from the
calorimeters and muon detectors to select the most interesting events in a fixed time interval of less than 4\mus. The high-level
 trigger processor farm further decreases the event rate from around 100\unit{kHz} to less than 1\unit{kHz}, before data storage.
A more detailed description of the CMS detector, together with a definition of the coordinate system used and the relevant
kinematic variables, can be found in Ref.~\cite{JINST}.
\section{Simulated samples}
\label{sec:mc}

The Monte Carlo (MC) simulation is used to design the
search, to aid in the estimation of SM  backgrounds, and to evaluate the sensitivity
to top squark pair production.

The \MGvATNLO2.2.2 generator~\cite{Alwall:2014hca}
in the leading-order (LO) mode, with MLM
matching~\cite{Alwall:2007fs}, and
with the LO NNPDF3.0~\cite{Ball:2014uwa} parton distribution functions (PDFs)
is used to generate top squark signal events as well as SM \ttbar, \wjets,
\zjets, and \gjets.
Single top quark events
 are generated at next-to-leading order (NLO) with \POWHEG
 2.0~\cite{Nason:2004rx,Frixione:2007vw,Alioli:2010xd,Re:2010bp}, while rare SM processes such as
$\ttbar\PZ$~and $\ttbar\PW$~are generated at NLO using the
\MGvATNLO2.2.2 program, with FxFx
matching~\cite{Frederix:2012ps} and the
NLO NNPDF3.0 PDFs. Parton showering, hadronization, and the underlying event are modeled by \PYTHIA 8.205~\cite{Sjostrand:2014zea}.
For SM processes, the response of the CMS detector is simulated with the \GEANTfour~\cite{geant4} package, while the CMS fast simulation program~\cite{fastsim} is used for the signal samples.
The most precise cross section calculations are used to normalize the SM simulated samples, corresponding most often to next-to-next-to-leading order (NNLO) accuracy.

To improve on the \MGvATNLO modeling of the
multiplicity of additional jets from initial state radiation (ISR),
simulated \ttbar events are reweighted based on the
number of ISR jets ($N_\mathrm{J}^\mathrm{ISR}$) so as to make the jet
multiplicity agree with data.
The same reweighting procedure is applied to SUSY MC events.
The reweighting factors vary between 0.92 and 0.51 for
$N_\mathrm{J}^\mathrm{ISR}$ between 1 and 6.  We take one half of the deviation
from unity as the systematic uncertainty on these reweighting factors to 
cover possible differences between top quark and top squark pair production.

\section{Event reconstruction and preselection}
\label{sec:evtsel}

Data events are selected online using triggers that require either
a large \pt imbalance or the presence of an isolated
electron or muon, see Table~\ref{tab:sels}.
The combined trigger efficiency, as measured with a data sample of events with large scalar sum of jet \pt, is
$>$99\% in the signal regions of interest described below.

The offline event reconstruction is based on the particle-flow (PF)
algorithm~\cite{Sirunyan:2017ulk}, which combines information
from the
tracker, calorimeter, and muon systems to identify charged and neutral hadrons, photons, electrons, and muons in the event.
The preselection based on PF objects is summarized in
Table~\ref{tab:sels}
and is described in more detail below.

The reconstructed vertex with the largest value of summed physics-object $\pt^2$ is taken to be the primary $\Pp\Pp$ interaction vertex. The physics objects are the objects returned by a jet finding algorithm~\cite{antikt,FastJet} applied to all charged tracks associated with the vertex, plus the corresponding associated missing transverse momentum.

Selected events are required to have exactly one electron~\cite{Khachatryan:2015hwa} or muon~\cite{MUOART} with $\pt>20\GeV$ and $\abs{\eta}<1.4442$ or $\abs{\eta}<2.4$, respectively.
The lepton needs to be consistent with originating from the
primary interaction vertex and isolated from other activity in the event.
Typical lepton selection efficiencies
are approximately 85\% for electrons and 95\% for muons  within the selection acceptance criteria, with
variations at the level of a few percent depending on the \pt and $\eta$ of the lepton.

Jets are formed by clustering neutral and charged PF objects using the
anti-\kt algorithm~\cite{antikt} with a distance parameter of 0.4.
The charged PF objects are required to be consistent with originating
from the primary vertex.
Jet energies are
corrected for contributions from multiple interactions in the same or adjacent
beam crossings
(pileup)~\cite{FastJet}, and to account for nonuniformity
in the detector response~\cite{Khachatryan:2016kdb}.
Jets overlapping with the selected lepton within a cone $\Delta R=\sqrt{\smash[b]{(\Delta\eta)^{2}+(\Delta\phi)^{2}}}=0.4$ are not considered.
We select events with two or more jets 
with $\pt>30\GeV$ and $\abs{\eta}<2.4$,
at least one of which is required to be consistent with containing the decay of a heavy-flavor hadron.
These jets, referred to as b-tagged jets, are identified using two
different working points (medium and tight WP)
of the CSVv2 tagging algorithm~\cite{ref:btag,CMS:2016kkf}.
The jet corrections described above are propagated consistently as a
correction to the missing transverse momentum vector (\ptvecmiss),
defined as the negative vector \pt sum of all PF
objects.  We denote the magnitude of this vector as $\ETmiss$ in the
discussion below.
Events with possible contributions from beam halo processes or anomalous noise in the calorimeter are rejected using dedicated filters~\cite{Chatrchyan:2011tn}.

\begin{table}[!htpb]
  \setlength{\extrarowheight}{.7em}
  \begin{center}
    \topcaption{\label{tab:sels} Summary of the event
      preselection.             The symbol
      $\pt^{\mathrm{lep}}$ denotes the \pt of the
      lepton, while $\pt^{\text{sum}}$ is the scalar \pt sum
      of PF candidates in a cone around the lepton but excluding
      the lepton.  For veto tracks this variable is calculated using
      charged PF candidates, while in the case of selected and veto
      leptons neutral PF candidates are also included. 
      The veto lepton and track definitions are used for event rejection as described in 
      the text.
      Light-flavor jets are defined as jets originating from u, d, s quarks or gluons.}
      \begin{tabular}{ l | l }
      \hline
      \multirow{2}{*}{Trigger} & $\ETmiss>120\GeV$ and $\HTmiss=\abs{\sum(\ptvec^{\text{\,jets}})+\ptvec^{\mathrm{^{\text{\,lep}}}}}>120\GeV$ or \\
      & isolated electron (muon):       $\pt^{\mathrm{lep}} > 25 (22)\GeV$, $\abs{\eta} < 2.1 (2.4)$ \\
      \hline
      Selected lepton & electron (muon): $\pt^{\mathrm{lep}} > 20\GeV$, $\abs{\eta} <1.442 (2.4)$ \\
      Selected lepton isolation & $\pt^{\text{sum}} < 0.1 \times
      \pt^{\mathrm{lep}}$,  $\Delta R = \min[0.2,\max(0.05, 10\GeV / \pt^{\mathrm{lep}})]$ \\
      Jets and b-tagged jets & $\pt > 30\GeV$, $\abs{\eta} < 2.4$ \\
      \ \ \ b tagging efficiency & medium (tight) WP: 60--70 (35--50)\% for jet \pt 30--400\GeV \\
      \ \ \ b tagging mistag rate & medium (tight) WP : $\sim1$\% ($\sim$0.2\%) for light-flavor quarks\\
            Missing transverse momentum & $\ETmiss > 250\GeV$ and $\Delta\phi\left(\ETmiss,J_{{1,2}}\right) > 0.8$ \\
      Transverse mass & $\MT > 150\GeV$ \\
      Veto lepton &  muon or electron with $\pt^{\mathrm{lep}} > 5\GeV$, $\abs{\eta} <2.4$ and \\
       & $\pt^{\text{sum}} < 0.1 \times \pt^{\text{lep}}$,  $\Delta R = \min[0.2,\max(0.05, 10\GeV / \pt^{\text{lep}})]$ \\
      Veto track &  charged PF candidate, $\pt > 10\GeV$, $\abs{\eta} <2.4$ and \\
       & $\pt^{\text{sum}} < \min\left(0.1 \times \pt^{\text{lep}}, 6\GeV\right)$, $\Delta R = 0.3$ \\
      \hline
    \end{tabular}
  \end{center}
\end{table}

Background events originating from \ttbar decays with only one top quark decaying leptonically
($\ttbar\to 1\ell$), \wjets, and single top quark processes are suppressed by the requirement on the 
\MET and the transverse mass (\MT) of the lepton-\ptvecmiss system. 
For signal, higher values of \MET than for background are expected due to the presence 
of additional unobserved particles, the LSPs. 
Similarly, the \MT
distribution has a jacobian edge around the \PW \xspace boson mass for background events, 
whereas for signal events no such edge exists due to the presence of the LSPs. 
We require \MT to be greater than 150\GeV. 
After these requirements, the largest contribution of SM background events is from processes with two lepton in the final state such as from \ttbar ($\ttbar\to2\ell$)
where the second lepton does not pass the selection requirements for the leading lepton.
Additional rejection is achieved by vetoing events containing a second lepton or isolated track passing looser identification
and isolation requirements than those used for the leading lepton. 
We also demand that the angle \minDPhiMETjet in
the azimuthal plane between the \ptvecmiss and the direction
of the closest of the two leading \pt  jets in the event
($J_1$ and $J_2$) to be greater than 0.8\unit{radians}.  This requirement is
motivated by the fact that
background $\ttbar\to 2\ell$ events tend to have high-\pt top quarks, and thus objects in these events tend to be collinear in the transverse plane,  resulting in smaller values of
\minDPhiMETjet than is typical for signal events.

\section{Signal regions}
\label{sec:srdef}

We define two sets of signal regions.  The first set (``standard'') is
designed to be sensitive to most of the
$\Delta m\left(\PSQt,\PSGczDo\right) \equiv   m_{\PSQt}-m_{\PSGczDo}$
parameter space, where $m_{\PSQt}$ and
$m_{\PSGczDo}$ are the masses of the top squark and the LSP, respectively.

The second set (``compressed'') is designed to enhance sensitivity
to the  decay mode in Fig.~\ref{fig:diagram}(a) when $\Delta m\left(\PSQt,\PSGczDo\right)\sim m_{\PQt}$.
While the signal regions within each set are mutually exclusive,
there is overlap across the signal regions of the two  sets.

Both sets have been optimized to have a high signal sensitivity for different decay modes 
and mass hypotheses using simulation of the SM background processes and the simplified model topologies 
shown in Fig.~\ref{fig:diagram}.

For the first set, signal regions are defined by
categorizing events based on the number of jets ($N_\mathrm{J}$),
the \MET, the invariant mass (\Mlb) of the lepton and the
closest b-tagged jet in $\Delta R$,
and a modified version of the topness variable~\cite{Graesser:2012qy}, \tmod:
\begin{equation}
\tmod = \ln(\min S),
\text{ with }
S(\vec{p}_{\PW}, p_{z},\nu) = \frac{(m_{\PW}^2-(p_\nu+p_{\ell})^2)^2}{a_{\PW}^4} +
\frac{(m_{\PQt}^2 - (p_{\PQb}+p_{\PW})^2)^2}{a_{\PQt}^4}
\label{eq:tmod}
\end{equation}
with the constraint $\ptvecmiss=\vec{p}_{\mathrm{T},\PW}+\vec{p}_{\mathrm{T},\nu}$.
The first term corresponds to the leptonically decaying top quark, the second term to the hadronically decaying top quark.
The calculation uses resolution parameters $a_{\PW} = 5\GeV$ and $a_{\PQt} = 15\GeV$.
The exact choices of objects used in this variable together with a more detailed motivation can be found in Ref.~\cite{Sirunyan:2016jpr}.

In models with \PSQt decays containing a \PSGcpmDo that is almost mass degenerate with the \PSGczDo, the SM decay products of the \PSGcpmDo are very soft. 
The final state for these signal can contain a small number of jets, while in signal models without this mass degeneracy at least four jets are expected. 
The \Mlb distribution has a sharp endpoint at about $\sqrt{m_{\PQt}^2-m_{\PW}^2}$ for events containing a leptonically decaying top quark such as \ttbar events or signals containing
at least one top quark in the decay chain. 
On the other hand, the \Mlb distribution does not have this endpoint for the sub-dominant background of \wjets{} as well as signal models with top squark decays to a b quark and a \PSGcpmDo. 
The \tmod variable tests for compatibility
with the $\ttbar\to 2\ell$ hypothesis when one of the leptons
is not reconstructed. 
Very high values of \tmod imply that an event is not compatible with the  $\ttbar\to 2\ell$ hypothesis. Signal models with large $\Delta m\left(\PSQt,\PSGczDo\right)$ result in such values.
On the other hand negative values of \tmod 
are a property of $\ttbar\to 2\ell$. As signal models with a small mass splitting between \PSQt and \PSGczDo 
also have low values in \tmod, we keep events with negative \tmod, to retain sensitivity for these signal models.

The requirements for the standard signal regions are summarized in Table~\ref{tab:SR}.

\begin{table}[hbt]
\topcaption{\protect Definitions for the 27 signal regions of the standard selection.
  At least one b-tagged jet satisfying the medium WP algorithm is required in all search regions.
  To suppress the \wjets\ background in signal regions with
  $\Mlb>175\GeV$, we instead use the more strict tight WP requirement.}
\label{tab:SR}
\centering
\begin{tabular}{crc|lllll}
\hline
$N_\mathrm{J}$ & \tmod & \multicolumn{1}{c}{\Mlb [\GeVns{}]} & \multicolumn{5}{c}{\MET [\GeVns{}]} \\
\hline
2--3 & $>$10 & $\leq175$     & 250-350, & 350-450, & 450-600, & $>$600 & \\
2--3 & $>$10 & $>$175        & 250-450, & 450-600, & $>$600 & & \\
$\geq$4 & $\leq0$ & $\leq175$   & 250-350, & 350-450, & 450-550, & 550-650, & $>$650 \\
$\geq$4 & $\leq0$ & $>$175      & 250-350, & 350-450, & 450-550, & $>$550 & \\
$\geq$4 & 0-10 & $\leq175$      & 250-350, & 350-550, & $>$550 & & \\
$\geq$4 & 0-10 & $>$175         & 250-450, & $>$450 & & & \\
$\geq$4 & $>$10 & $\leq175$     & 250-350, & 350-450, & 450-600, & $>$600 & \\
$\geq$4 & $>$10 & $>$175        & 250-450, & $>$450 & & & \\
\hline
\end{tabular}
\end{table}

The compressed signal regions are designed to
select events with
a high-\pt jet from ISR, which is needed to provide the necessary boost to the system to
obtain large \ETmiss and large \MT.  Thus, we require at least five jets in the event,
with the highest \pt jet failing the medium WP of the b tagging algorithm.
Additionally, we reject events if the selected lepton has $\pt > 150\GeV$ as we expect the lepton to be soft in the compressed region.
We also require the angle between the lepton direction and \ptvecmiss in the azimuthal plane to be $<$2.  This is because the ISR selection results in boosted top squarks
with decay products typically close to each other.
Finally, we relax the \minDPhiMETjet requirement in the preselection from 0.8 to 0.5 to increase the signal acceptance. 
The selection requirements for the compressed signal regions are summarized in Table~\ref{tab:compSR}.

\begin{table}[hbt]
\setlength{\extrarowheight}{.7em}
\topcaption{Summary of the compressed selection and the requirements for the
  four corresponding signal regions.  The symbol
$\Delta\phi(\ETmiss,\ell) $ denotes the angle between \ptvecmiss and
the $\vec{\pt}$ of the lepton, and $J_1$ denotes the highest \pt jet.}
\label{tab:compSR}
\centering
\begin{tabular}{l|lllll}
\hline
 \multirow{2}{*}{Selection} & $\NJ \geq5$, & \multicolumn{2}{l}{$J_1$ not b tagged,} &  \multicolumn{2}{l}{$\Delta\phi(\ETmiss,\ell) <2$,} \\
 & \multicolumn{3}{l}{$\minDPhiMETjet >0.5$,} & \multicolumn{2}{l}{$\pt^{\ell} <150\GeV$} \\
 \hline
 Search regions & \ETmiss= & \!\!\!250-350, & \!\!\!350-450, & \!\!\!450-550, & \!\!\!$>550\GeV$ \\
\hline
\end{tabular}
\end{table}

\section{Background estimation}
\label{Sec:BkgEst}

Three categories of background from SM processes remain after the selection requirements described in Sections~\ref{sec:evtsel} and~\ref{sec:srdef}.

\begin{itemize}
  \item Lost-lepton background: events with two leptonically decaying
    \PW\ bosons
in which one of the leptons is not reconstructed
    or identified.  This background arises primarily from \ttbar\ events, with a smaller contribution from single-top quark processes.
    It is the dominant background in the $\Mlb < 175\GeV$ and $\NJ \geq 4$ search regions, and is estimated using a dilepton control sample.
  \item One-lepton background: events with a single leptonically decaying \PW\ boson and no additional source of genuine \ETmiss.
 This background is strongly suppressed by the preselection requirements of $\MET>250\GeV$ and $\MT>150\GeV$.

    The suppression is much more effective for events with a \PW\ boson originating from the decay of a top quark than for direct \PW\ boson production (\wjets),
    as the mass of the top quark imposes a bound on the mass of the charged lepton-neutrino system.
    As a result, the tail of the \MT distribution in $\ttbar\to 1\ell$ events is dominated by \MET resolution effects,
    while in \wjets\ it extends further and is largely driven by the width of the \PW\ boson.

    The \wjets\ background estimate is obtained from a control sample of events with no b-tagged jets.
    The subleading $\ttbar\to 1\ell$  background is modeled from simulation.
    One-lepton events are the dominant background in the $\Mlb \geq 175\GeV$ search regions.
  \item $\PZ\to\nu\bar{\nu}$ background: events with exactly one leptonically decaying \PW\ boson and a \PZ\ boson that decays to a pair of neutrinos, e.g., $\ttbar \PZ$ or $\PW \PZ$.
    This background is estimated from simulation, after normalizing the simulated event yield to the observed data counts in a control region obtained by
    selecting events with three leptons, two of which must be consistent with the \PZ\ decay hypothesis.
\end{itemize}

These three types of backgrounds are discussed below.
More details about the validity of the background estimation methods for the two first categories can be found in Ref.~\cite{Sirunyan:2016jpr}.

\subsection{Lost-lepton background}\label{sec:dilepton}

The lost-lepton background is estimated from a dilepton control sample obtained with the same selection requirements as the signal
sample, except for requiring the presence of a second isolated lepton with $\pt>10\GeV$.
 For each signal region, a corresponding control region is
 constructed, with an exception
as noted below.
In defining the control regions, 
the $\vec{\pt}$ of the second lepton is added to the \ptvecmiss and all relevant event quantities are recalculated. 
The estimated background in each search region is then obtained from the yield of data events in the
control region and a transfer factor defined as the ratio of the expected SM event yields in the signal and control regions, as determined from simulation.
Corrections obtained from studies of $\ensuremath{\cPZ/\Pgg^\star} \to \ell \ell$ events are applied
to account for small
differences in lepton reconstruction and selection efficiencies between data and simulation.

\begin{figure}[thb]
\begin{center}
\includegraphics[width=0.667\textwidth]{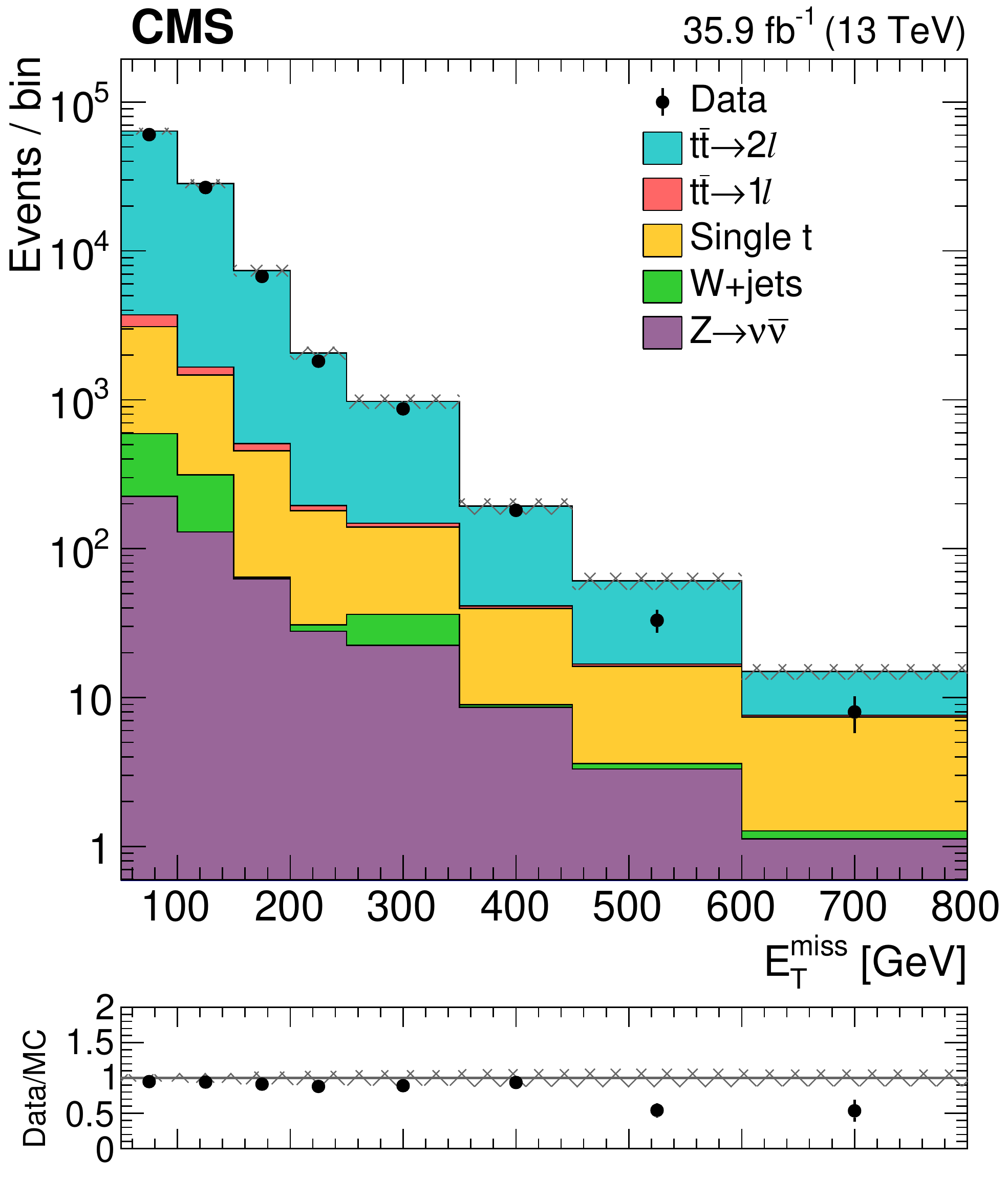}
\end{center}
\caption{\label{fig:2lmet} Distributions in \MET for a top quark enriched control region of \Pe\Pgm\ events with at least one b-tagged jet.
The ratio of data to simulation as a function of \MET is also shown. 
It is taken as a correction of the \MET distribution in simulation of \ttbar and $\PQt\PW$ events with a lost lepton.
}
\end{figure}

Due to a lack of statistics, the two or three highest \ETmiss bins of Table~\ref{tab:SR}
are combined resulting in the list of control regions listed in Table~\ref{tab:CR2L},
and the simulation, after the correction described below, is used
to determine the expected distribution of SM events as a function of \MET.
The correction is based on a study of the \MET distribution in a top quark enriched control region
of \Pe\Pgm\ events with at least one b-tagged jet, as shown in Fig.~\ref{fig:2lmet}.
The ratio of data to simulation yields as a function of \MET in the \Pe\Pgm\ sample
is taken as a bin-by-bin correction for the expected \MET distribution
in the simulation of \ttbar and $\PQt\PW$ events with a lost lepton.
The uncertainty in each bin is taken to be one half the deviation from unity.

\begin{table}[hbt]
    \topcaption{Dilepton control regions utilizing combined bins in \MET.}  \label{tab:CR2L}
\centering
\begin{tabular}{crc|lll}
\hline
$N_\mathrm{j}$ & \tmod & \multicolumn{1}{c}{\Mlb [\GeVns{}]} & \multicolumn{3}{c}{\MET [\GeVns{}]} \\
\hline
  2--3 & $>$10 & $>$175 & 250-450, & 450-600, & $>$600 \\
  $\geq$4 & 0--10 & $\leq$175 & 350-550, & $>$550 & \\
  $\geq$4 & 0--10 & $>$175 & 250-450, & $>$450 & \\
  $\geq$4 & $>$10 & $>$175 & 250-450, & $>$450 & \\
  \hline
\end{tabular}
\end{table}

The dominant uncertainties on the transfer factors arise from the statistical uncertainties in the simulated samples
and the uncertainties in the lepton efficiency.  These range from 5--100\% and 5--15\%, respectively. 
The uncertainties on the lepton efficiency are derived from studies of 
samples of leptonically-decaying \PZ{} bosons.
For the regions of Table~\ref{tab:CR2L}, there are also uncertainties associated with
the \ETmiss distribution.  These are also dominated by the statistical precision of the simulated samples, and range between 10 and 100\%.
Uncertainties due to the jet energy scale and the b tagging efficiency 
are evaluated by varying the correction factors for simulation by their uncertainties,
and the uncertainties due to the choices of renormalization and factorization scale used in the generation of SM samples are assessed by varying the scales by a factor of 2.
All these uncertainties are found to be small.
The resulting systematic uncertainties on the transfer factors are 10--100\%, depending on the region.
These are generally smaller than the statistical uncertainties from the data yield in the corresponding
control regions that
are used, in conjunction with the transfer factors, to predict the SM background in the signal regions.

\subsection{One-lepton background}
\label{sec:onelepton}

As discussed previously,
the one-lepton background receives contributions from processes where the leptonically decaying
\PW\ boson is produced directly or from the decay of a top quark.
The background from direct \PW\ boson production is estimated in each search region using
a control region obtained with the same selection as the signal region except that the b tagging requirement is inverted to enrich the sample in \wjets\ events.
The estimate in each search region is then obtained using a transfer factor determined from simulated samples that accounts for the b quark jet acceptance
and tagging efficiency. The estimate is corrected for small differences in the performance of the b tagging algorithm between data and simulation.

\begin{figure}[!htpb]
\centering
      \includegraphics[width=0.95\textwidth]{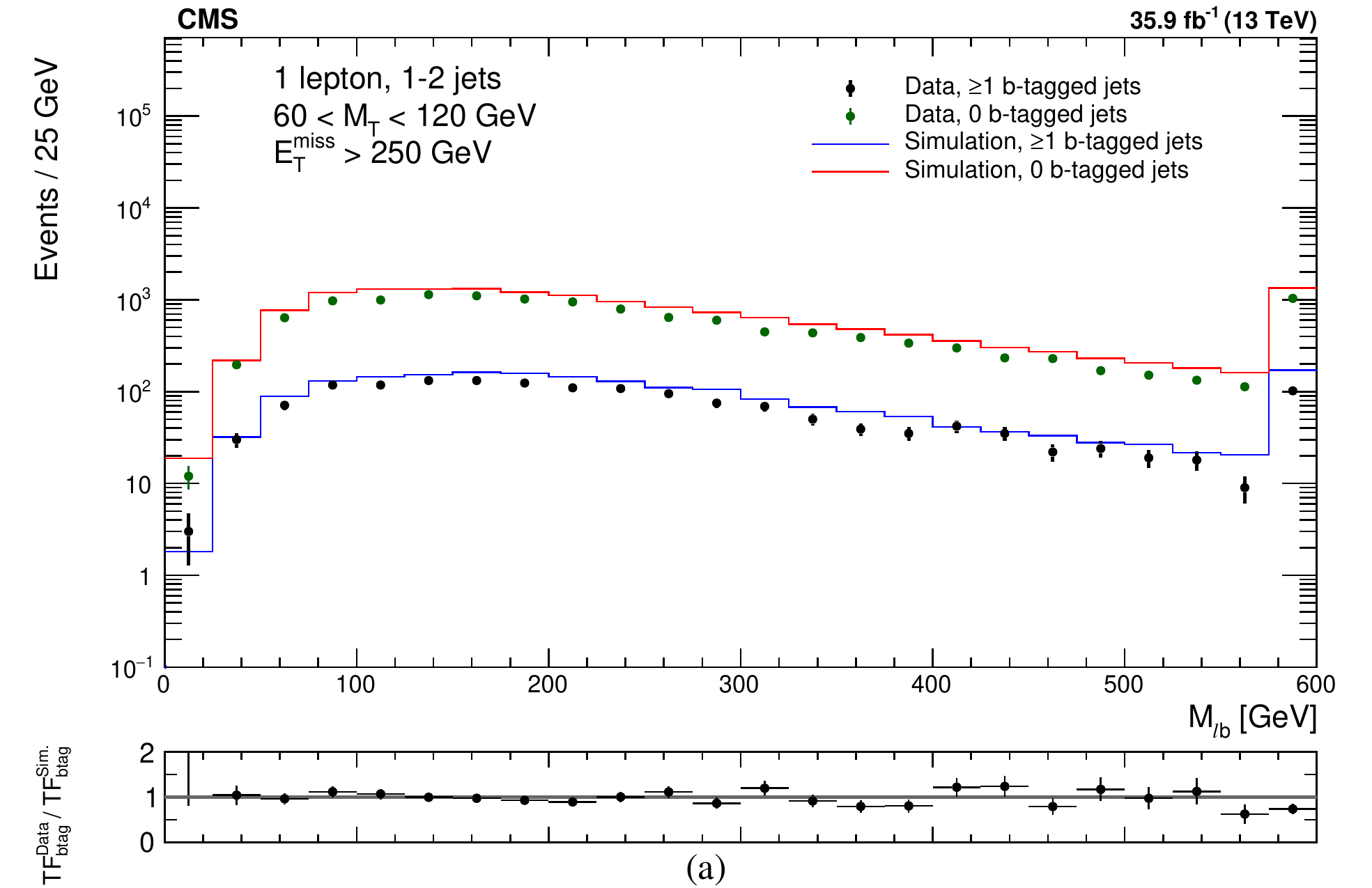}\vskip 0.5cm
        \includegraphics[width=0.45\textwidth]{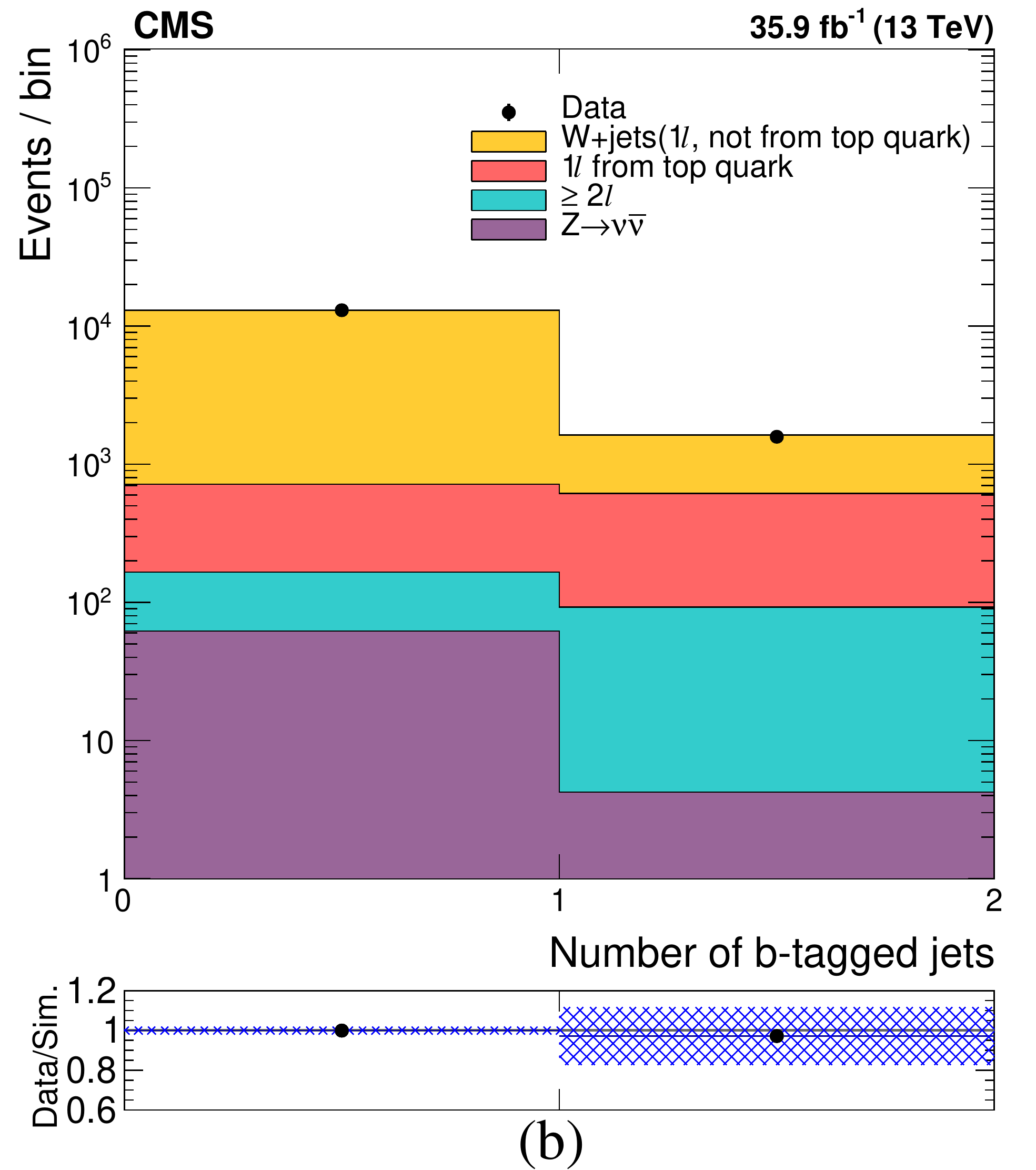}        \includegraphics[width=0.45\textwidth]{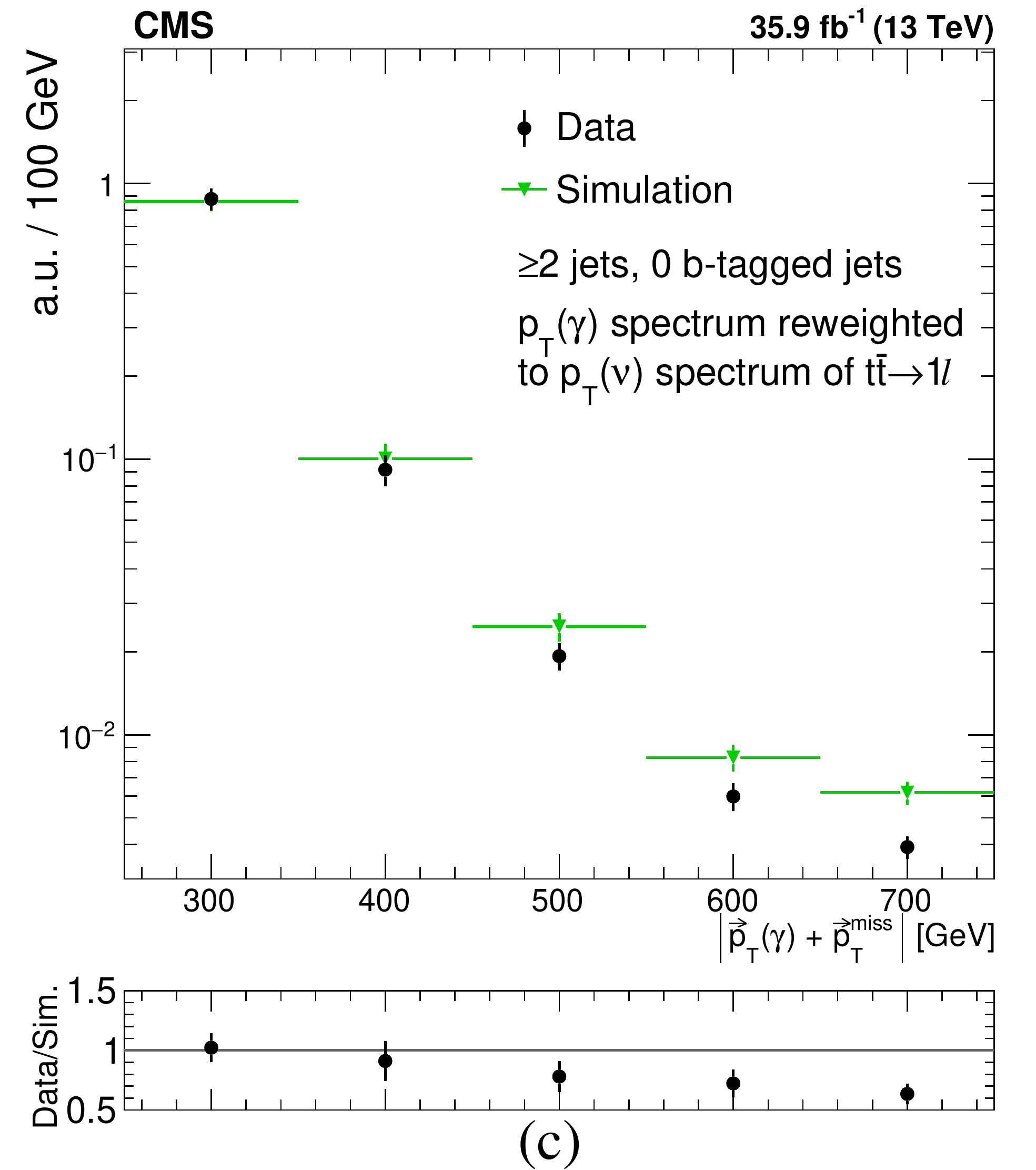}        \vspace{5.5pt}
\caption{\label{fig:onelep} Comparison of the modeling of kinematic distributions in data and simulation relevant for the estimate of the single lepton backgrounds.
  (a)  Distribution in \Mlb in a control sample with 1 or 2 jets, with $60 < \MT < 120\GeV$ and $\MET > 250\GeV$.
  The distribution is shown separately for events with 0 and $\geq$1 jet passing the requirement of the medium b tagging WP.  The lower panel shows the ratio of the transfer factors (TF) from the 0 b-tagged jets to the $\geq$1 b-tagged jets samples, in data and simulation. The uncertainty shown is statistical only.
  (b) Distribution in the number of b-tagged jets in the same control sample.   The shaded band shows the uncertainty resulting from a 50\% systematic uncertainty in the heavy flavor component of the \wjets\ sample.
  (c) Comparison of the \MET distribution between data and simulation in the \gjets\ control region. The uncertainty shown is statistical only. The ratio between data and simulation shown on the lower panel is used to correct the simulation for its \MET resolution.
}
\end{figure}

In the control sample, the \Mlb variable is constructed using the selected lepton and the jet in the event with the highest value of the
b-tag discriminator.  The \Mlb distribution is validated in a control sample enriched in the \wjets\ events, obtained by selecting events with 1 or 2 jets,
$60 < \MT < 120\GeV$, $\MET>250\GeV$, and either 0 or $\geq 1$ jet passing the medium WP of the b tagging algorithm.  
Figure~\ref{fig:onelep}(a) shows the \Mlb distribution in both data and simulation for the control samples with 0 and $\geq$1 b-tagged jets. The bottom panel shows 
the good agreement between data and simulation in the extrapolation factor from the 0 b-tagged jets sample to the sample with $\geq$1 b-tagged jets. 

The largest uncertainty in the transfer factor comes from the limited event counts of the simulated samples, followed by the uncertainty on the heavy-flavor fraction of jets in \wjets\ events.
A comparison of the multiplicity of b-tagged jets between data and
simulation is performed in a \wjets\ enriched region obtained with the
same selection as for the \Mlb distribution, as shown in
Fig.~\ref{fig:onelep}(b).  The difference between data and simulation is covered by a 50\% uncertainty on the heavy-flavor component of \wjets\ events, and is
indicated by the shaded band in the figure.
Variations of the jet energy scale and b tagging efficiency within their measured uncertainties each result in a 10\% uncertainty in the background estimate.
The total uncertainty in the estimate of the \wjets\ background varies from 20 to 80\%, depending on signal region.

Simulation studies indicate that in all signal regions the contribution from $\ttbar\to 1\ell$ events is expected to be smaller than 10\% of the total background.
This estimate is sensitive to the correct modeling of the \MET resolution, since this affects the \MT tail.
The modeling of the \MET resolution is studied using data and simulated
samples of  \gjets\ events.
The photon \pt spectrum is reweighted to match that of the neutrino in simulated $\ttbar\to 1\ell$
events after first re-weighting
the photon \pt spectrum in simulation to match that observed in data.
We then add the photon $\vec{\pt}$ to the \MET, and compare the resulting spectra.
Differences of up to 40\% in the \MET shape between data and simulated events are observed, as shown in Fig.~\ref{fig:onelep}(c) for a selection with at least 2 jets.
Corrections for these differences are applied to the $\ttbar\to 1\ell$ simulation and a resulting 100\% uncertainty is assigned to the estimate of this background.

\subsection{Background from events containing \texorpdfstring{$\PZ\to\nu\bar{\nu}$}{ZtoNuNu} decays}

The third and last category of background arises from $\ttbar \PZ$, \PW\PZ, and other rare multiboson processes,
all with a leptonically decaying \PW\ boson and one or more \PZ\ bosons decaying to neutrinos.
Within this category, the contribution from \PW\PZ\ events is dominant in the low-\NJ\ bins, whereas in events with higher \NJ, 60--80\% of this background is due to $\ttbar\PZ$ processes.

The background from these processes is estimated from simulation with normalization obtained from a data control sample containing three leptons.
For this sample, two leptons must form an opposite charge, same flavor pair having an invariant mass between 76 and 106 GeV.
The normalization of the $\PW\PZ$ and $\ttbar\PZ$ processes is determined by performing a template fit to the distribution of the number of b-tagged jets in this sample.
The result of this fit yield scale factors of $1.21 \pm0.11$ and $1.14 \pm 0.30$ to be applied to the simulated samples of $\PW\PZ$ and $\ttbar\PZ$ events, respectively.

We also assess all relevant theoretical and experimental uncertainties that can affect the shapes of the kinematic distributions of our signal region definitions by 
recomputing acceptances after modifying the various kinematical quantities and
reconstruction efficiencies within
their respective uncertainties.
The experimental uncertainties are obtained by variations of the simulation correction factors within their measured uncertainties.
The largest contributions are due to the uncertainties in the jet energy scale and to the choices of the renormalization and factorization scales used
in the MC generation of SM samples. 
The latter is obtained by varying the scales by a factor of 2.
Other uncertainties are due to the lepton and b tagging efficiencies, the modeling of additional jets in the parton shower, pileup, the value of the strong coupling constant $\alpS$, and the PDF sets.
The uncertainty on the PDF sets is evaluated by using replicas of the NNPDF3.0 set~\cite{Ball:2014uwa}.

The total uncertainty in the $\PZ\to\nu\bar{\nu}$ background is 17--78\%, depending on the search region.

\section{Results and interpretation}
\label{sec:results}

\begin{figure}[htb]
\centering
\includegraphics[width=0.90\textwidth]{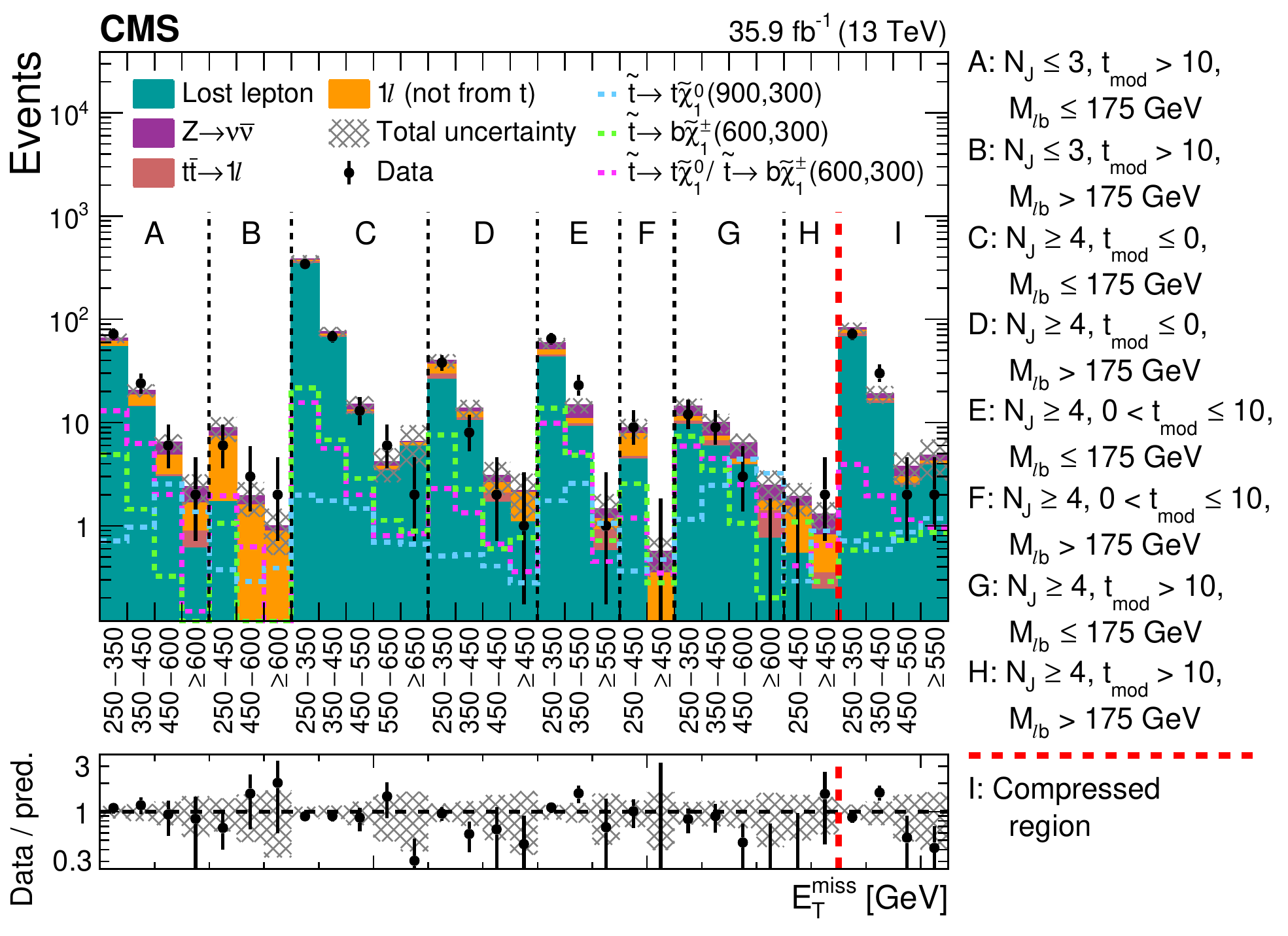}
\caption{\label{fig:results} Observed data yields compared with the SM background estimations for the 31 signal regions of Tables~\ref{tab:SR} and~\ref{tab:compSR}.
The total uncertainty in the background estimate, determined as the sum in quadrature of statistical and systematic uncertainties, is shown as a shaded band.
The expectations for three signal hypotheses are overlaid.
The corresponding numbers in parentheses in the legend refer to the
masses in~\GeV of the top squark and the neutralino. }
\end{figure}

The event yields in data in the 31 search regions
defined in Tables~\ref{tab:SR} and~\ref{tab:compSR}
 are statistically compatible with the estimated backgrounds from SM
 processes.  They are summarized in Table~\ref{tab:results} and
 Fig.~\ref{fig:results} and are
interpreted in the context of the simplified models of top squark pair production described in Section~\ref{sec:intro}.
 Further information
on the experimental results to facilitate reinterpretations for beyond the SM models not considered here is given in Appendix A.

\begin{table}[htb]
\centering
\topcaption{\label{tab:results}Result of the background estimates and
data yields corresponding to \Lint, for the 31 signal regions of Tables~\ref{tab:SR} and~\ref{tab:compSR}.}
\resizebox{\textwidth}{!}{\begin{tabular}{cccccccccc}
\hline
 \multirow{2}{*}{$\NJ$} & \multirow{2}{*}{$\tmod$} & $\Mlb$ & $\MET$ & Lost  & \multirow{2}{*}{1$\ell$ (top)} & 1$\ell$ (not & \multirow{2}{*}{$Z\to\nu\bar{\nu}$} & Total & \multirow{2}{*}{Data} \\
  &  &  [\GeVns{}] &  [\GeVns{}] &  lepton &  &  top) &  & background &  \\
\hline
$\leq3$ &    $>$10 & $\leq175$ & 250--350 & 53.9$\pm$6.2 & $<0.1$ & 7.2$\pm$2.5 & 4.7$\pm$1.2 & 65.8$\pm$6.8 & 72 \\
$\leq3$ &    $>$10 & $\leq175$ & 350--450 & 14.2$\pm$2.4 & 0.2$\pm$0.2 & 4.1$\pm$1.4 & 2.1$\pm$0.8 & 20.5$\pm$2.9 & 24 \\
$\leq3$ &    $>$10 & $\leq175$ & 450--600 & 2.9$\pm$0.9 & 0.1$\pm$0.1 & 1.7$\pm$0.7 & 1.6$\pm$0.5 & 6.4$\pm$1.3 & 6 \\
$\leq3$ &    $>$10 & $\leq175$ &    $>$600 & 0.6$\pm$0.5 & 0.3$\pm$0.3 & 0.8$\pm$0.3 & 0.7$\pm$0.4 & 2.4$\pm$0.8 & 2 \\
\hline
$\leq3$ &    $>$10 &     $>$175 & 250--450 & 1.7$\pm$0.8 & $<0.1$ & 5.6$\pm$2.2 & 1.5$\pm$0.5 & 8.9$\pm$2.4 & 6 \\
$\leq3$ &    $>$10 &     $>$175 & 450--600 & 0.02$\pm$0.01 & $<0.1$ & 1.6$\pm$0.6 & 0.4$\pm$0.3 & 1.9$\pm$0.7 & 3 \\
$\leq3$ &    $>$10 &     $>$175 &    $>$600 & 0.01$\pm$0.01 & $<0.1$ & 0.9$\pm$0.4 & 0.1$\pm$0.3 & 1.0$\pm$0.5 & 2 \\
\hline
$\geq$4 & $\leq0$ & $\leq175$ & 250--350 & 346$\pm$30 & 13.2$\pm$13.2 & 9.7$\pm$8.6 & 14.4$\pm$3.9 & 383$\pm$34 & 343 \\
$\geq$4 & $\leq0$ & $\leq175$ & 350--450 & 66.3$\pm$7.9 & 2.3$\pm$2.3 & 2.5$\pm$1.7 & 4.4$\pm$1.2 & 75.5$\pm$8.5 & 68 \\
$\geq$4 & $\leq0$ & $\leq175$ & 450--550 & 12.1$\pm$2.8 & 0.6$\pm$0.6 & 0.5$\pm$0.5 & 1.8$\pm$0.5 & 15.0$\pm$2.9 & 13 \\
$\geq$4 & $\leq0$ & $\leq175$ & 550--650 & 3.4$\pm$1.5 & 0.1$\pm$0.1 & 0.3$\pm$0.2 & 0.4$\pm$0.1 & 4.1$\pm$1.5 & 6 \\
$\geq$4 & $\leq0$ & $\leq175$ &    $>$650 & 5.9$\pm$2.8 & $<0.1$ & 0.4$\pm$0.4 & 0.2$\pm$0.1 & 6.6$\pm$2.9 & 2 \\
\hline
$\geq$4 & $\leq0$ &     $>$175 & 250--350 & 26.0$\pm$4.3 & 3.1$\pm$3.1 & 7.5$\pm$3.0 & 3.0$\pm$0.9 & 39.7$\pm$6.2 & 38 \\
$\geq$4 & $\leq0$ &     $>$175 & 350--450 & 10.4$\pm$2.6 & 0.6$\pm$0.6 & 1.6$\pm$0.7 & 1.2$\pm$0.4 & 13.7$\pm$2.8 & 8 \\
$\geq$4 & $\leq0$ &     $>$175 & 450--550 & 1.7$\pm$0.9 & 0.4$\pm$0.4 & 0.6$\pm$0.3 & 0.5$\pm$0.2 & 3.1$\pm$1.1 & 2 \\
$\geq$4 & $\leq0$ &     $>$175 &    $>$550 & 1.1$\pm$0.8 & $<0.1$ & 1.0$\pm$0.6 & 0.09$\pm$0.03 & 2.2$\pm$1.0 & 1 \\
\hline
$\geq$4 &   0--10 & $\leq175$ & 250--350 & 43.0$\pm$5.9 & 1.7$\pm$1.7 & 5.7$\pm$3.0 & 8.3$\pm$2.2 & 58.7$\pm$7.2 & 65 \\
$\geq$4 &   0--10 & $\leq175$ & 350--550 & 9.1$\pm$2.0 & 0.5$\pm$0.5 & 1.2$\pm$0.5 & 3.9$\pm$1.1 & 14.7$\pm$2.4 & 23 \\
$\geq$4 &   0--10 & $\leq175$ &    $>$550 & 0.6$\pm$0.3 & 0.3$\pm$0.3 & 0.3$\pm$0.2 & 0.3$\pm$0.3 & 1.5$\pm$0.6 & 1 \\
\hline
$\geq$4 &   0--10 &     $>$175 & 250--450 & 4.4$\pm$1.4 & 0.3$\pm$0.3 & 3.1$\pm$1.3 & 1.1$\pm$0.3 & 8.9$\pm$1.9 & 9 \\
$\geq$4 &   0--10 &     $>$175 &    $>$450 & 0.10$\pm$0.17 & $<0.1$ & 0.2$\pm$0.3 & 0.2$\pm$0.1 & 0.6$\pm$0.2 & 0 \\
\hline
$\geq$4 &    $>$10 & $\leq175$ & 250--350 & 9.5$\pm$2.3 & 0.8$\pm$0.8 & 1.1$\pm$0.9 & 3.0$\pm$0.8 & 14.3$\pm$2.7 & 12 \\
$\geq$4 &    $>$10 & $\leq175$ & 350--450 & 5.9$\pm$1.8 & 0.7$\pm$0.7 & 0.7$\pm$0.5 & 2.7$\pm$0.8 & 10.0$\pm$2.1 & 9 \\
$\geq$4 &    $>$10 & $\leq175$ & 450--600 & 3.8$\pm$1.3 & 0.1$\pm$0.1 & 0.4$\pm$0.3 & 2.0$\pm$0.5 & 6.3$\pm$1.5 & 3 \\
$\geq$4 &    $>$10 & $\leq175$ &    $>$600 & 0.8$\pm$0.6 & 0.7$\pm$0.7 & 0.3$\pm$0.4 & 0.7$\pm$0.3 & 2.4$\pm$1.0 & 0 \\
\hline
$\geq$4 &    $>$10 &     $>$175 & 250--450 & 0.5$\pm$0.3 & $<0.1$ & 1.0$\pm$0.6 & 0.4$\pm$0.1 & 1.9$\pm$0.7 & 0 \\
$\geq$4 &    $>$10 &     $>$175 &    $>$450 & 0.2$\pm$0.2 & 0.1$\pm$0.1 & 0.5$\pm$0.3 & 0.5$\pm$0.2 & 1.3$\pm$0.4 & 2 \\
\hline
\hline
\multicolumn{3}{l|}{Compressed region} & 250--350 & 67.5$\pm$8.9 & 5.3$\pm$5.3 & 5.0$\pm$1.8 & 4.3$\pm$1.2 & 82$\pm$11 & 72 \\
\multicolumn{3}{l|}{Compressed region} & 350--450 & 15.1$\pm$3.5 & 1.0$\pm$1.0 & 0.8$\pm$0.3 & 1.9$\pm$0.6 & 18.9$\pm$3.7 & 30 \\
\multicolumn{3}{l|}{Compressed region} & 450--550 & 2.4$\pm$1.3 & 0.1$\pm$0.1 & 0.4$\pm$0.2 & 0.8$\pm$0.3 & 3.7$\pm$1.4 & 2 \\
\multicolumn{3}{l|}{Compressed region} &    $>$550 & 3.9$\pm$2.0 & 0.1$\pm$0.1 & 0.2$\pm$0.2 & 0.6$\pm$0.2 & 4.8$\pm$2.0 & 2 \\
\hline
\end{tabular}
}
\end{table}

\begin{figure}[htb]
\centering
\includegraphics[width=0.8\textwidth]{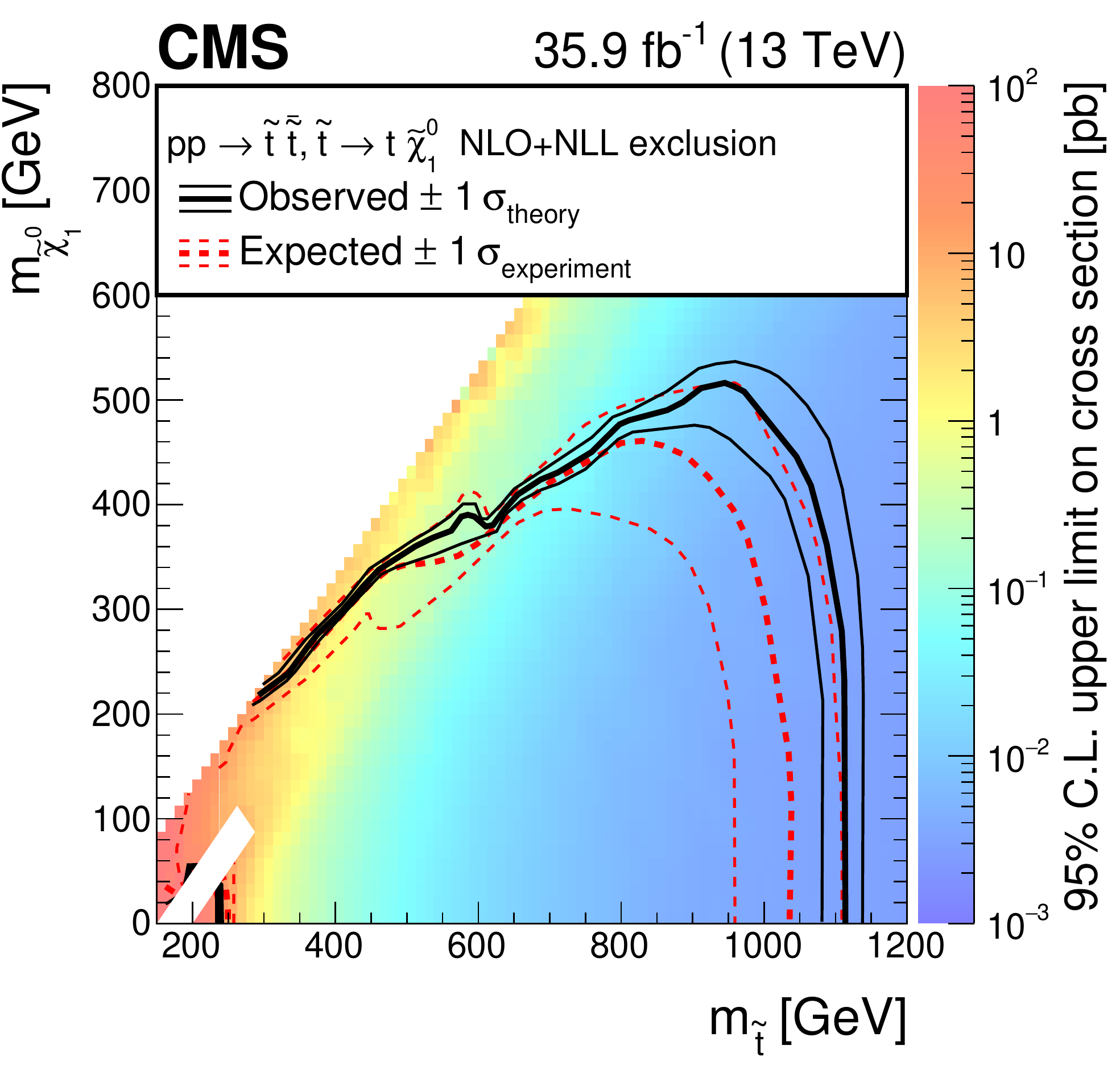}
\caption{\label{fig:limits:T2tt}The exclusion limits at 95\% CL for
  direct top squark pair production with decay
$\PSQt\,\PASQt  \to \PQt\PSGczDo ~ \PAQt\PSGczDo $.
 The interpretation is done in the two-dimensional space of $m_{\PSQt}$ vs. $m_{\PSGczDo}$. The color indicates the 95\% CL upper limit
  on the cross section times branching fraction at each point in the $m_{\PSQt}$ vs. $m_{\PSGczDo}$ plane.
  The area below the thick black curve represents
  the observed exclusion region at 95\% CL
assuming 100\% branching fraction,
while the dashed red lines indicate the expected limits at 95\% CL and their ${\pm}1\sigma$ experimental standard deviation uncertainties.  The thin black lines show the effect of the theoretical uncertainties ($\sigma_\text{theory}$) in the signal cross section. The whited out region is discussed in Section~\ref{sec:results}.}
\end{figure}

\begin{figure}[htb]
\centering
\includegraphics[width=0.8\textwidth]{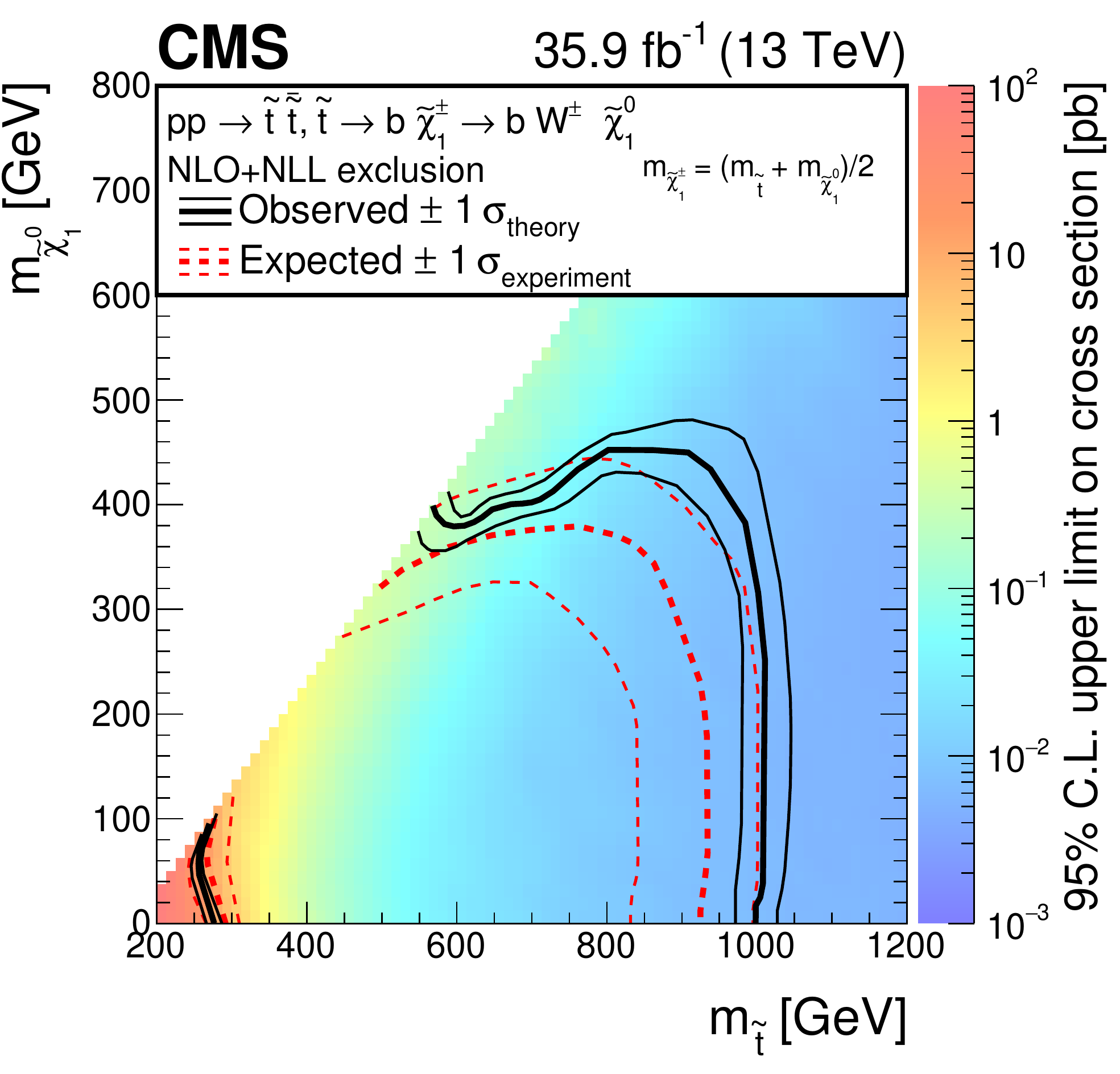}
\caption{\label{fig:limits:T2bW}The exclusion limit at 95\% CL for
  direct top squark pair production
  with decay
   $\PSQt\,\PASQt\to \PQb \PSGcpDo \PAQb \PSGcmDo$,
  $\PSGcpmDo\to\PW^{\pm}\PSGczDo$. The mass of the chargino is chosen to be $(m_{\PSQt} + m_{\PSGczDo})/2$.
  The interpretation is done in the two-dimensional space of $m_{\PSQt}$ vs. $m_{\PSGczDo}$. The color indicates the 95\% CL upper limit
  on the cross section times branching fraction at each point in the $m_{\PSQt}$ vs. $m_{\PSGczDo}$ plane.
  The area between the thick black curves represents the observed
  exclusion region at 95\% CL
assuming 100\% branching fraction,
while the dashed red lines indicate the expected limits at 95\% CL and their ${\pm}1\sigma$ experimental standard deviation uncertainties.  The thin black lines show the effect of the theoretical uncertainties ($\sigma_\text{theory}$) in the signal cross section.}
\end{figure}

\begin{figure}[htb]
\centering
\includegraphics[width=0.8\textwidth]{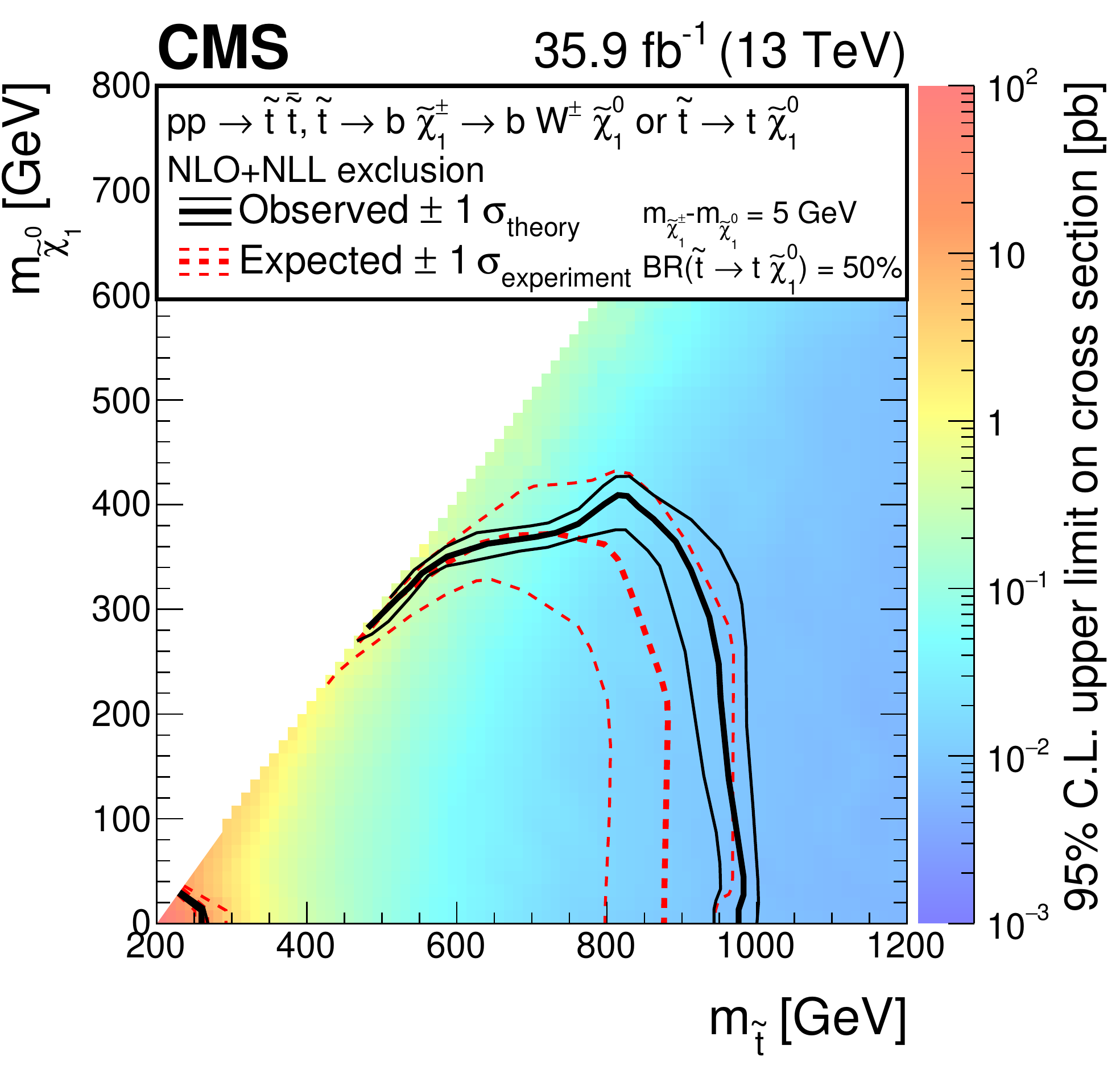}
\caption{\label{fig:limits:T2tb}The exclusion limit at 95\% CL for
  direct top squark pair production
  with decay
  $\PSQt\,\PASQt \to \PQb \PSGcpDo \PAQt \PSGczDo$,
   $\PSGcpmDo\to\PW^{\pm}\PSGczDo$.
  The mass splitting of the chargino and neutralino is fixed to 5\GeV.
 The interpretation is done in the two-dimensional space of $m_{\PSQt}$ vs. $m_{\PSGczDo}$. The color indicates the 95\% CL upper limit
  on the cross section at each point in the $m_{\PSQt}$ vs. $m_{\PSGczDo}$ plane.
  The area between the thick black curves represents the observed exclusion region at 95\% CL, while the dashed red lines indicate the expected limits at 95\% CL and their ${\pm}1\sigma$ experimental standard deviation uncertainties.  The thin black lines show the effect of the theoretical uncertainties ($\sigma_\text{theory}$) in the signal cross section.}
\end{figure}

For a given model, limits on the production cross-section are derived as a function of the masses of the SUSY particles by combining search regions using a modified frequentist approach, employing the $\mathrm{CL}_\mathrm{s}$ criterion in an asymptotic formulation~\cite{Junk:1999kv,Read:2002hq,Cowan:2010jsX,LHC-HCG}.  These limits are turned into exclusion regions in the $m(\PSQt)-m(\PSGczDo)$ plane using the calculation of the cross-section from reference ~\cite{Borschensky:2014cia} and are shown on Figures ~\ref{fig:limits:T2tt}, ~\ref{fig:limits:T2bW}, and~\ref{fig:limits:T2tb}.
Limits are obtained by combining the 27 regions from the standard selection defined in Table~\ref{tab:SR},
except for the model of Fig.~\ref{fig:diagram}(a) in the compressed region
$100 \leq \Delta m\left(\PSQt,\PSGczDo\right) \leq 225\GeV$, where
we use the four
compressed search regions listed in Table~\ref{tab:compSR}.
This approach
improves the expected cross section upper limit in the compressed mass region by  $\sim$15--30\%.
When computing the limits, the expected signal yields are corrected for possible contamination of SUSY events
in the data control regions.  These corrections are typically
around 5--10\%.

A summary of the uncertainties in the signal efficiency is shown in Table~\ref{tab:syst}. 
They are evaluated in the same manner as done in the background estimation methods, described in Section~\ref{Sec:BkgEst}.
The largest uncertainties are due to the limited size of the simulated signal samples, the b tagging efficiency, and the jet energy scale.
For model points with a small mass splitting, the ISR uncertainty described in Section~\ref{sec:mc} is also significant.
Since new physics signals are simulated using the CMS fast simulation program, additional uncertainties are assigned
to the correction of the lepton and b tagging efficiencies, as well as
to cover differences in \MET resolution between the fast simulation and the full \GEANTfour-based model of the CMS detector.
The latter uncertainty is small in the bulk of the model space, but may reach up to 25\% in scenarios with a compressed mass spectrum.
Uncertainties due to the integrated luminosity, ISR modeling, \ETmiss\
resolution, and b tagging and lepton efficiencies are treated as fully
correlated across search regions.

\begin{table}[htb]
\centering
\topcaption{\label{tab:syst} Summary of the systematic uncertainties in the signal efficiency.}
\begin{tabular}{lc}\hline
Source & Typical range of values [\%] \\
\hline
Simulation statistical uncertainty & 5--25 \\
Renormalization and factorization scales & 2--4 \\
Integrated luminosity & 2.5 \\
Trigger & 2--4 \\
b tagging scale factors & 1--7 \\
Jet energy scale & 1--20 \\
Lepton identification and veto efficiency & 1--4 \\
ISR modeling & 2--15 \\
\MET modeling & 1--10 \\
\hline
Total uncertainty & 7--38 \\
\hline
\end{tabular}
\end{table}

Figure~\ref{fig:limits:T2tt} shows the 95\% confidence level (CL) upper limit on
$\Pp \Pp \to \PSQt\,\PASQt \to \PQt \PSGczDo~\PAQt \PSGczDo$,
assuming unpolarized top quarks in the decay chain, together with the upper limit at 95\% CL on the signal cross section.
We exclude top squark masses up to 1120\GeV for a massless LSP and LSP masses up to 515\GeV for a 950\GeV top squark mass.
The white band corresponds to the region $|m_{\PSQt}-m_{\PQt}-m_{\PSGczDo}| < 25\GeV$, $m_{\PSQt}<275\GeV$
where the selection efficiency of top squark events changes rapidly and becomes very sensitive to details of the model and the simulation.  No cross section limit is established in that region.

Figure~\ref{fig:limits:T2bW} shows the 95\% CL upper limit for $\Pp\Pp\to\PSQt\,\PASQt\to \PQb\PAQb\PSGcpmDo\PSGcpmDo$, $\PSGcpmDo\to\PW\PSGczDo$, together with the upper limit at 95\% CL on the excluded signal cross section. The mass of the chargino is chosen to be $(m_{\PSQt} + m_{\PSGczDo})/2$.
We exclude top squark masses up to 1000\GeV for a massless LSP and LSP masses up to 450\GeV for a 800\GeV top squark mass.

Figure~\ref{fig:limits:T2tb} shows the 95\% CL upper limit for $\Pp\Pp\to\PSQt\,\PASQt\to\PQt\PQb\PSGcpmDo\PSGczDo$, $\PSGcpmDo\to\PW^{*}\PSGczDo$, together with the upper limit at 95\% CL on the excluded signal cross section. The mass splitting of the chargino and neutralino is fixed to 5\GeV.
We exclude top squark masses up to 980\GeV for a massless LSP and LSP masses up to 400\GeV for a 825\GeV top squark mass.

\section{Summary}
We have reported on a search for top squark pair production in \pp collisions at $\sqrt{s}=13\TeV$ in events with a single isolated electron or muon, jets, and large missing transverse momentum using data collected with the CMS detector during the 2016 run of the LHC, corresponding to an integrated luminosity of \Lint.
The event data yields are consistent with the expectations from SM processes.
The results are interpreted as exclusion limits in the context of supersymmetric models with pair production of top squarks
that decay either to a top quark and a neutralino or to a bottom quark and a chargino.
Assuming both top squarks decay to a top quark and a neutralino, we exclude at 95\% CL top squark masses up to 1120\GeV
for a massless neutralino and neutralino masses up to 515\GeV for a 950\GeV top squark mass.
For a scenario where both top squarks decay to a bottom quark and a chargino, with the chargino mass the average of the masses of the neutralino and top squark,
we exclude at the 95\% CL top squark masses up to 1000\GeV for a massless neutralino and neutralino masses up to 450\GeV for a 800\GeV top squark mass.
For the mixed decay scenario, with the mass splitting between the chargino and neutralino fixed to be 5\GeV,
we exclude at the 95\% CL top squark masses up to 980\GeV for a massless neutralino and neutralino masses up to 400\GeV for a 825\GeV top squark mass.
\begin{acknowledgments}
We congratulate our colleagues in the CERN accelerator departments for the excellent performance of the LHC and thank the technical and administrative staffs at CERN and at other CMS institutes for their contributions to the success of the CMS effort. In addition, we gratefully acknowledge the computing centres and personnel of the Worldwide LHC Computing Grid for delivering so effectively the computing infrastructure essential to our analyses. Finally, we acknowledge the enduring support for the construction and operation of the LHC and the CMS detector provided by the following funding agencies: BMWFW and FWF (Austria); FNRS and FWO (Belgium); CNPq, CAPES, FAPERJ, and FAPESP (Brazil); MES (Bulgaria); CERN; CAS, MoST, and NSFC (China); COLCIENCIAS (Colombia); MSES and CSF (Croatia); RPF (Cyprus); SENESCYT (Ecuador); MoER, ERC IUT, and ERDF (Estonia); Academy of Finland, MEC, and HIP (Finland); CEA and CNRS/IN2P3 (France); BMBF, DFG, and HGF (Germany); GSRT (Greece); OTKA and NIH (Hungary); DAE and DST (India); IPM (Iran); SFI (Ireland); INFN (Italy); MSIP and NRF (Republic of Korea); LAS (Lithuania); MOE and UM (Malaysia); BUAP, CINVESTAV, CONACYT, LNS, SEP, and UASLP-FAI (Mexico); MBIE (New Zealand); PAEC (Pakistan); MSHE and NSC (Poland); FCT (Portugal); JINR (Dubna); MON, RosAtom, RAS, RFBR and RAEP (Russia); MESTD (Serbia); SEIDI, CPAN, PCTI and FEDER (Spain); Swiss Funding Agencies (Switzerland); MST (Taipei); ThEPCenter, IPST, STAR, and NSTDA (Thailand); TUBITAK and TAEK (Turkey); NASU and SFFR (Ukraine); STFC (United Kingdom); DOE and NSF (USA).
\end{acknowledgments}

\clearpage
\bibliography{auto_generated}

\providecommand{\href}[2]{#2}\begingroup\raggedright\begin{thebibliography}{10}%
\makeatletter
\providecommand{\hrefCMSnoop }[0]{\@secondoftwo}%
\makeatother
\providecommand{\doi}{\texttt{doi:}\begingroup \urlstyle{tt}\Url}

\bibitem{Ramond:1971gb}
\hrefCMSnoop {}{P.~Ramond, ``{Dual theory for free fermions}'',} \textit{ Phys.
  Rev. D} \textbf{ 3} (1971) 2415,
\href{http://dx.doi.org/10.1103/PhysRevD.3.2415}{\doi{10.1103/PhysRevD.3.2415}}.

\bibitem{Golfand:1971iw}
\hrefCMSnoop {}{Y.~A. Golfand and E.~P. Likhtman, ``{Extension of the algebra
  of {P}oincar\'{e} group generators and violation of {P} invariance}'',}
  \textit{ JETP Lett.} \textbf{ 13} (1971)
323.

\bibitem{Neveu:1971rx}
\hrefCMSnoop {}{A.~Neveu and J.~H. Schwarz, ``{Factorizable dual model of
  pions}'',} \textit{ Nucl. Phys. B} \textbf{ 31} (1971) 86,
\href{http://dx.doi.org/10.1016/0550-3213(71)90448-2}{\doi{10.1016/0550-3213(71)90448-2}}.

\bibitem{Volkov:1972jx}
\hrefCMSnoop {}{D.~V. Volkov and V.~P. Akulov, ``{Possible universal neutrino
  interaction}'',} \textit{ JETP Lett.} \textbf{ 16} (1972)
438.

\bibitem{Wess:1973kz}
\hrefCMSnoop {}{J.~Wess and B.~Zumino, ``{A {L}agrangian model invariant under
  supergauge transformations}'',} \textit{ Phys. Lett. B} \textbf{ 49} (1974)
  52,
\href{http://dx.doi.org/10.1016/0370-2693(74)90578-4}{\doi{10.1016/0370-2693(74)90578-4}}.

\bibitem{Wess:1974tw}
\hrefCMSnoop {}{J.~Wess and B.~Zumino, ``{Supergauge transformations in four
  dimensions}'',} \textit{ Nucl. Phys. B} \textbf{ 70} (1974) 39,
\href{http://dx.doi.org/10.1016/0550-3213(74)90355-1}{\doi{10.1016/0550-3213(74)90355-1}}.

\bibitem{Fayet:1974pd}
\hrefCMSnoop {}{P.~Fayet, ``{Supergauge invariant extension of the {H}iggs
  mechanism and a model for the electron and its neutrino}'',} \textit{ Nucl.
  Phys. B} \textbf{ 90} (1975) 104,
\href{http://dx.doi.org/10.1016/0550-3213(75)90636-7}{\doi{10.1016/0550-3213(75)90636-7}}.

\bibitem{Nilles:1983ge}
\hrefCMSnoop {}{H.~P. Nilles, ``{Supersymmetry, supergravity and particle
  physics}'',} \textit{ Phys. Rep.} \textbf{ 110} (1984) 1,
\href{http://dx.doi.org/10.1016/0370-1573(84)90008-5}{\doi{10.1016/0370-1573(84)90008-5}}.

\bibitem{Aad:2012tfa}
\hrefCMSnoop {}{{ATLAS Collaboration}, ``{Observation of a new particle in the
  search for the Standard Model Higgs boson with the ATLAS detector at the
  LHC}'',} \textit{ Phys. Lett. B} \textbf{ 716} (2012) 1,
  \href{http://dx.doi.org/10.1016/j.physletb.2012.08.020}{\doi{10.1016/j.physletb.2012.08.020}},
\href{http://www.arXiv.org/abs/1207.7214}{\texttt{arXiv:1207.7214}}.

\bibitem{Chatrchyan:2012xdj}
\hrefCMSnoop {}{{CMS Collaboration}, ``{Observation of a new boson at a mass of
  125 GeV with the CMS experiment at the LHC}'',} \textit{ Phys. Lett. B}
  \textbf{ 716} (2012) 30,
  \href{http://dx.doi.org/10.1016/j.physletb.2012.08.021}{\doi{10.1016/j.physletb.2012.08.021}},
\href{http://www.arXiv.org/abs/1207.7235}{\texttt{arXiv:1207.7235}}.

\bibitem{Chatrchyan:2012tx}
\hrefCMSnoop {}{{CMS Collaboration}, ``{Combined results of searches for the
  standard model Higgs boson in pp collisions at $\sqrt{s}=7$ TeV}'',} \textit{
  Phys. Lett. B} \textbf{ 710} (2012) 26,
  \href{http://dx.doi.org/10.1016/j.physletb.2012.02.064}{\doi{10.1016/j.physletb.2012.02.064}},
\href{http://www.arXiv.org/abs/1202.1488}{\texttt{arXiv:1202.1488}}.

\bibitem{Papucci:2011wy}
\hrefCMSnoop {}{M.~Papucci, J.~T. Ruderman, and A.~Weiler, ``{Natural {SUSY}
  endures}'',} \textit{ JHEP} \textbf{ 09} (2012) 035,
  \href{http://dx.doi.org/10.1007/JHEP09(2012)035}{\doi{10.1007/JHEP09(2012)035}},
\href{http://www.arXiv.org/abs/1110.6926}{\texttt{arXiv:1110.6926}}.

\bibitem{Barbieri:2009ev}
\hrefCMSnoop {}{R.~Barbieri and D.~Pappadopulo, ``{S-particles at their
  naturalness limits}'',} \textit{ JHEP} \textbf{ 10} (2009) 061,
  \href{http://dx.doi.org/10.1088/1126-6708/2009/10/061}{\doi{10.1088/1126-6708/2009/10/061}},
\href{http://www.arXiv.org/abs/0906.4546}{\texttt{arXiv:0906.4546}}.

\bibitem{Dimopoulos:1995mi}
\hrefCMSnoop {}{S.~Dimopoulos and G.~F. Giudice, ``{Naturalness constraints in
  supersymmetric theories with nonuniversal soft terms}'',} \textit{ Phys.
  Lett. B} \textbf{ 357} (1995) 573,
  \href{http://dx.doi.org/10.1016/0370-2693(95)00961-J}{\doi{10.1016/0370-2693(95)00961-J}},
\href{http://www.arXiv.org/abs/hep-ph/9507282}{\texttt{arXiv:hep-ph/9507282}}.

\bibitem{ATLASstop1L2015}
\hrefCMSnoop {}{{ATLAS Collaboration}, ``{Search for top squarks in final
  states with one isolated lepton, jets, and missing transverse momentum in
  $\sqrt{s}=13$ TeV $pp$ collisions with the ATLAS detector}'',} \textit{ Phys.
  Rev. D} \textbf{ 94} (2016) 052009,
  \href{http://dx.doi.org/10.1103/PhysRevD.94.052009}{\doi{10.1103/PhysRevD.94.052009}},
\href{http://www.arXiv.org/abs/1606.03903}{\texttt{arXiv:1606.03903}}.

\bibitem{Sirunyan:2016jpr}
\hrefCMSnoop {}{{CMS Collaboration}, ``{Searches for pair production of
  third-generation squarks in $\sqrt{s}=13$ $\,\text {TeV}$ pp collisions}'',}
  \textit{ Eur. Phys. J. C} \textbf{ 77} (2017) 327,
  \href{http://dx.doi.org/10.1140/epjc/s10052-017-4853-2}{\doi{10.1140/epjc/s10052-017-4853-2}},
\href{http://www.arXiv.org/abs/1612.03877}{\texttt{arXiv:1612.03877}}.

\bibitem{Khachatryan:2017rhw}
\hrefCMSnoop {}{{CMS Collaboration}, ``Search for supersymmetry in the
  all-hadronic final state using top quark tagging in $pp$ collisions at
  {$\sqrt{s} = 13$\TeV}'',} \textit{ Phys. Rev. D} \textbf{ 96} (2017) 012004,
  \href{http://dx.doi.org/10.1103/PhysRevD.96.012004}{\doi{10.1103/PhysRevD.96.012004}},
\href{http://www.arXiv.org/abs/1701.01954}{\texttt{arXiv:1701.01954}}.

\bibitem{JINST}
\hrefCMSnoop {}{{CMS Collaboration}, ``The {CMS} experiment at the {CERN}
  {LHC}'',} \textit{ JINST} \textbf{ 3} (2008) S08004,
\href{http://dx.doi.org/10.1088/1748-0221/3/08/S08004}{\doi{10.1088/1748-0221/3/08/S08004}}.

\bibitem{ArkaniHamed:2007fw}
N.~Arkani-Hamed\hrefCMSnoop {}{ {et~al.}, ``{MARMOSET:} the path from {LHC}
  data to the new standard model via on-shell effective theories'',} (2007).
\href{http://www.arXiv.org/abs/hep-ph/0703088}{\texttt{arXiv:hep-ph/0703088}}.

\bibitem{Alwall:2008ag}
\hrefCMSnoop {}{J.~Alwall, P.~Schuster, and N.~Toro, ``Simplified models for a
  first characterization of new physics at the {LHC}'',} \textit{ Phys. Rev. D}
  \textbf{ 79} (2009) 075020,
  \href{http://dx.doi.org/10.1103/PhysRevD.79.075020}{\doi{10.1103/PhysRevD.79.075020}},
\href{http://www.arXiv.org/abs/0810.3921}{\texttt{arXiv:0810.3921}}.

\bibitem{Alwall:2008va}
\hrefCMSnoop {}{J.~Alwall, M.-P. Le, M.~Lisanti, and J.~G. Wacker,
  ``Model-independent jets plus missing energy searches'',} \textit{ Phys. Rev.
  D} \textbf{ 79} (2009) 015005,
  \href{http://dx.doi.org/10.1103/PhysRevD.79.015005}{\doi{10.1103/PhysRevD.79.015005}},
\href{http://www.arXiv.org/abs/0809.3264}{\texttt{arXiv:0809.3264}}.

\bibitem{Alves:2011wf}
\hrefCMSnoop {}{{LHC New Physics Working Group} Collaboration, ``Simplified
  models for {LHC} new physics searches'',} \textit{ J. Phys. G} \textbf{ 39}
  (2012) 105005,
  \href{http://dx.doi.org/10.1088/0954-3899/39/10/105005}{\doi{10.1088/0954-3899/39/10/105005}},
\href{http://www.arXiv.org/abs/1105.2838}{\texttt{arXiv:1105.2838}}.

\bibitem{Alwall:2014hca}
J.~Alwall\hrefCMSnoop {}{ {et~al.}, ``The automated computation of tree-level
  and next-to-leading order differential cross sections, and their matching to
  parton shower simulations'',} \textit{ JHEP} \textbf{ 07} (2014) 079,
  \href{http://dx.doi.org/10.1007/JHEP07(2014)079}{\doi{10.1007/JHEP07(2014)079}},
\href{http://www.arXiv.org/abs/1405.0301}{\texttt{arXiv:1405.0301}}.

\bibitem{Alwall:2007fs}
J.~Alwall\hrefCMSnoop {}{ {et~al.}, ``Comparative study of various algorithms
  for the merging of parton showers and matrix elements in hadronic
  collisions'',} \textit{ Eur. Phys. J. C} \textbf{ 53} (2008) 473,
  \href{http://dx.doi.org/10.1140/epjc/s10052-007-0490-5}{\doi{10.1140/epjc/s10052-007-0490-5}},
\href{http://www.arXiv.org/abs/0706.2569}{\texttt{arXiv:0706.2569}}.

\bibitem{Ball:2014uwa}
\hrefCMSnoop {}{{NNPDF} Collaboration, ``{Parton distributions for the LHC Run
  II}'',} \textit{ JHEP} \textbf{ 04} (2015) 040,
  \href{http://dx.doi.org/10.1007/JHEP04(2015)040}{\doi{10.1007/JHEP04(2015)040}},
\href{http://www.arXiv.org/abs/1410.8849}{\texttt{arXiv:1410.8849}}.

\bibitem{Nason:2004rx}
\hrefCMSnoop {}{P.~Nason, ``{A new method for combining NLO QCD with shower
  Monte Carlo algorithms}'',} \textit{ JHEP} \textbf{ 11} (2004) 040,
  \href{http://dx.doi.org/10.1088/1126-6708/2004/11/040}{\doi{10.1088/1126-6708/2004/11/040}},
\href{http://www.arXiv.org/abs/hep-ph/0409146}{\texttt{arXiv:hep-ph/0409146}}.

\bibitem{Frixione:2007vw}
\hrefCMSnoop {}{S.~Frixione, P.~Nason, and C.~Oleari, ``{Matching NLO QCD
  computations with parton shower simulations: the \POWHEG method}'',} \textit{
  JHEP} \textbf{ 11} (2007) 070,
  \href{http://dx.doi.org/10.1088/1126-6708/2007/11/070}{\doi{10.1088/1126-6708/2007/11/070}},
\href{http://www.arXiv.org/abs/0709.2092}{\texttt{arXiv:0709.2092}}.

\bibitem{Alioli:2010xd}
\hrefCMSnoop {}{S.~Alioli, P.~Nason, C.~Oleari, and E.~Re, ``{A general
  framework for implementing NLO calculations in shower Monte Carlo programs:
  the \POWHEG BOX}'',} \textit{ JHEP} \textbf{ 06} (2010) 043,
  \href{http://dx.doi.org/10.1007/JHEP06(2010)043}{\doi{10.1007/JHEP06(2010)043}},
\href{http://www.arXiv.org/abs/1002.2581}{\texttt{arXiv:1002.2581}}.

\bibitem{Re:2010bp}
\hrefCMSnoop {}{E.~Re, ``{Single-top $Wt$-channel production matched with
  parton showers using the POWHEG method}'',} \textit{ Eur. Phys. J. C}
  \textbf{ 71} (2011) 1547,
  \href{http://dx.doi.org/10.1140/epjc/s10052-011-1547-z}{\doi{10.1140/epjc/s10052-011-1547-z}},
\href{http://www.arXiv.org/abs/1009.2450}{\texttt{arXiv:1009.2450}}.

\bibitem{Frederix:2012ps}
\hrefCMSnoop {}{R.~Frederix and S.~Frixione, ``{Merging meets matching in
  MC@NLO}'',} \textit{ JHEP} \textbf{ 12} (2012) 061,
  \href{http://dx.doi.org/10.1007/JHEP12(2012)061}{\doi{10.1007/JHEP12(2012)061}},
\href{http://www.arXiv.org/abs/1209.6215}{\texttt{arXiv:1209.6215}}.

\bibitem{Sjostrand:2014zea}
T.~Sj{\"o}strand\hrefCMSnoop {}{ {et~al.}, ``An introduction to {PYTHIA}
  8.2'',} \textit{ Comput. Phys. Commun.} \textbf{ 191} (2015) 159,
  \href{http://dx.doi.org/10.1016/j.cpc.2015.01.024}{\doi{10.1016/j.cpc.2015.01.024}},
\href{http://www.arXiv.org/abs/1410.3012}{\texttt{arXiv:1410.3012}}.

\bibitem{geant4}
\hrefCMSnoop {}{{{GEANT4}} Collaboration, ``{GEANT4}---a simulation toolkit'',}
  \textit{ Nucl. Instrum. Meth. A} \textbf{ 506} (2003) 250,
\href{http://dx.doi.org/10.1016/S0168-9002(03)01368-8}{\doi{10.1016/S0168-9002(03)01368-8}}.

\bibitem{fastsim}
S.~Abdullin\hrefCMSnoop {}{ {et~al.}, ``The fast simulation of the {CMS}
  detector at {LHC}'',} \textit{ J. Phys. Conf. Ser.} \textbf{ 331} (2011)
  032049,
\href{http://dx.doi.org/10.1088/1742-6596/331/3/032049}{\doi{10.1088/1742-6596/331/3/032049}}.

\bibitem{Sirunyan:2017ulk}
\hrefCMSnoop {}{{CMS Collaboration}, ``{Particle-flow reconstruction and global
  event description with the CMS detector}'',} (2017).
  \href{http://www.arXiv.org/abs/1706.04965}{\texttt{arXiv:1706.04965}}.
Submitted to \textit{JINST}.

\bibitem{antikt}
\hrefCMSnoop {}{M.~Cacciari, G.~P. Salam, and G.~Soyez, ``The anti-$k_t$ jet
  clustering algorithm'',} \textit{ JHEP} \textbf{ 04} (2008) 063,
  \href{http://dx.doi.org/10.1088/1126-6708/2008/04/063}{\doi{10.1088/1126-6708/2008/04/063}},
  \href{http://www.arXiv.org/abs/0802.1189}{\texttt{arXiv:0802.1189}}.

\bibitem{FastJet}
\hrefCMSnoop {}{M.~Cacciari, G.~P. Salam, and G.~Soyez, ``{FastJet} user
  manual'',} \textit{ Eur. Phys. J. C} \textbf{ 72} (2012) 1896,
  \href{http://dx.doi.org/10.1140/epjc/s10052-012-1896-2}{\doi{10.1140/epjc/s10052-012-1896-2}},
\href{http://www.arXiv.org/abs/1111.6097}{\texttt{arXiv:1111.6097}}.

\bibitem{Khachatryan:2015hwa}
\hrefCMSnoop {}{{CMS Collaboration}, ``Performance of electron reconstruction
  and selection with the cms detector in proton-proton collisions at $\sqrt{s}
  = 8$~tev'',} \textit{ JINST} \textbf{ 10} (2015) P06005,
  \href{http://dx.doi.org/10.1088/1748-0221/10/06/P06005}{\doi{10.1088/1748-0221/10/06/P06005}},
\href{http://www.arXiv.org/abs/1502.02701}{\texttt{arXiv:1502.02701}}.

\bibitem{MUOART}
\hrefCMSnoop {}{{CMS Collaboration}, ``{Performance of CMS muon reconstruction
  in pp collision events at $\sqrt{s}=7$ TeV}'',} \textit{ JINST} \textbf{ 7}
  (2012) P10002,
  \href{http://dx.doi.org/10.1088/1748-0221/7/10/P10002}{\doi{10.1088/1748-0221/7/10/P10002}},
\href{http://www.arXiv.org/abs/1206.4071}{\texttt{arXiv:1206.4071}}.

\bibitem{Khachatryan:2016kdb}
\hrefCMSnoop {}{{CMS Collaboration}, ``{Jet energy scale and resolution in the
  CMS experiment in pp collisions at 8 TeV}'',} \textit{ JINST} \textbf{ 12}
  (2017) P02014,
  \href{http://dx.doi.org/10.1088/1748-0221/12/02/P02014}{\doi{10.1088/1748-0221/12/02/P02014}},
\href{http://www.arXiv.org/abs/1607.03663}{\texttt{arXiv:1607.03663}}.

\bibitem{ref:btag}
\hrefCMSnoop {}{{CMS Collaboration}, ``{Identification of b-quark jets with the
  CMS experiment}'',} \textit{ JINST} \textbf{ 8} (2013) P04013,
  \href{http://dx.doi.org/10.1088/1748-0221/8/04/P04013}{\doi{10.1088/1748-0221/8/04/P04013}},
\href{http://www.arXiv.org/abs/1211.4462}{\texttt{arXiv:1211.4462}}.

\bibitem{CMS:2016kkf}
\href {http://cds.cern.ch/record/2138504}{{{CMS}} Collaboration,
  ``Identification of b quark jets at the {CMS} experiment in the {LHC} run
  2'',} CMS Physics Analysis Summary CMS-PAS-BTV-15-001, 2016.

\bibitem{Chatrchyan:2011tn}
\hrefCMSnoop {}{{CMS Collaboration}, ``{Missing transverse energy performance
  of the CMS detector}'',} \textit{ JINST} \textbf{ 6} (2011) P09001,
  \href{http://dx.doi.org/10.1088/1748-0221/6/09/P09001}{\doi{10.1088/1748-0221/6/09/P09001}},
\href{http://www.arXiv.org/abs/1106.5048}{\texttt{arXiv:1106.5048}}.

\bibitem{Graesser:2012qy}
\hrefCMSnoop {}{M.~L. Graesser and J.~Shelton, ``Hunting mixed top squark
  decays'',} \textit{ Phys. Rev. Lett.} \textbf{ 111} (2013) 121802,
  \href{http://dx.doi.org/10.1103/PhysRevLett.111.121802}{\doi{10.1103/PhysRevLett.111.121802}},
\href{http://www.arXiv.org/abs/1212.4495}{\texttt{arXiv:1212.4495}}.

\bibitem{Junk:1999kv}
\hrefCMSnoop {}{T.~Junk, ``{Confidence level computation for combining searches
  with small statistics}'',} \textit{ Nucl. Instrum. Meth. A} \textbf{ 434}
  (1999) 435,
  \href{http://dx.doi.org/10.1016/S0168-9002(99)00498-2}{\doi{10.1016/S0168-9002(99)00498-2}},
\href{http://www.arXiv.org/abs/hep-ex/9902006}{\texttt{arXiv:hep-ex/9902006}}.

\bibitem{Read:2002hq}
\hrefCMSnoop {}{A.~L. Read, ``Presentation of search results: the {$CL_{S}$}
  technique'',} \textit{ J. Phys. G} \textbf{ 28} (2002) 2693,
  \href{http://dx.doi.org/10.1088/0954-3899/28/10/313}{\doi{10.1088/0954-3899/28/10/313}}.

\bibitem{Cowan:2010jsX}
\hrefCMSnoop {}{G.~Cowan, K.~Cranmer, E.~Gross, and O.~Vitells, ``Asymptotic
  formulae for likelihood-based tests of new physics'',} \textit{ Eur. Phys. J.
  C} \textbf{ 71} (2011) 1554,
  \href{http://dx.doi.org/10.1140/epjc/s10052-011-1554-0}{\doi{10.1140/epjc/s10052-011-1554-0}},
  \href{http://www.arXiv.org/abs/1007.1727}{\texttt{arXiv:1007.1727}}.
[Erratum: \DOI{10.1140/epjc/s10052-013-2501-z}].

\bibitem{LHC-HCG}
\href {http://cdsweb.cern.ch/record/1379837}{{ATLAS and CMS Collaborations, LHC
  Higgs Combination Group}, ``Procedure for the {LHC} {H}iggs boson search
  combination in {S}ummer 2011'',} Technical Report ATL-PHYS-PUB 2011-11, CMS
  NOTE 2011/005, 2011.

\bibitem{Borschensky:2014cia}
C.~Borschensky\hrefCMSnoop {}{ {et~al.}, ``{Squark and gluino production cross
  sections in $pp$ collisions at $\sqrt{s} = 13$, 14, 33 and 100 TeV}'',}
  \textit{ Eur. Phys. J. C} \textbf{ 74} (2014) 3174,
  \href{http://dx.doi.org/10.1140/epjc/s10052-014-3174-y}{\doi{10.1140/epjc/s10052-014-3174-y}},
\href{http://www.arXiv.org/abs/1407.5066}{\texttt{arXiv:1407.5066}}.

\bibitem{Collaboration:2242860}
\href {http://cds.cern.ch/record/2242860}{{CMS Collaboration}, ``{Simplified
  likelihood for the re-interpretation of public CMS results}'',} Technical
  Report CERN-CMS-NOTE-2017-001, CERN, Geneva, 2017.

\end{thebibliography}\endgroup
\clearpage
\appendix
\section{Additional information}
\label{app_outreach}

\begin{figure}[htb]
\centering
\includegraphics[width=\textwidth]{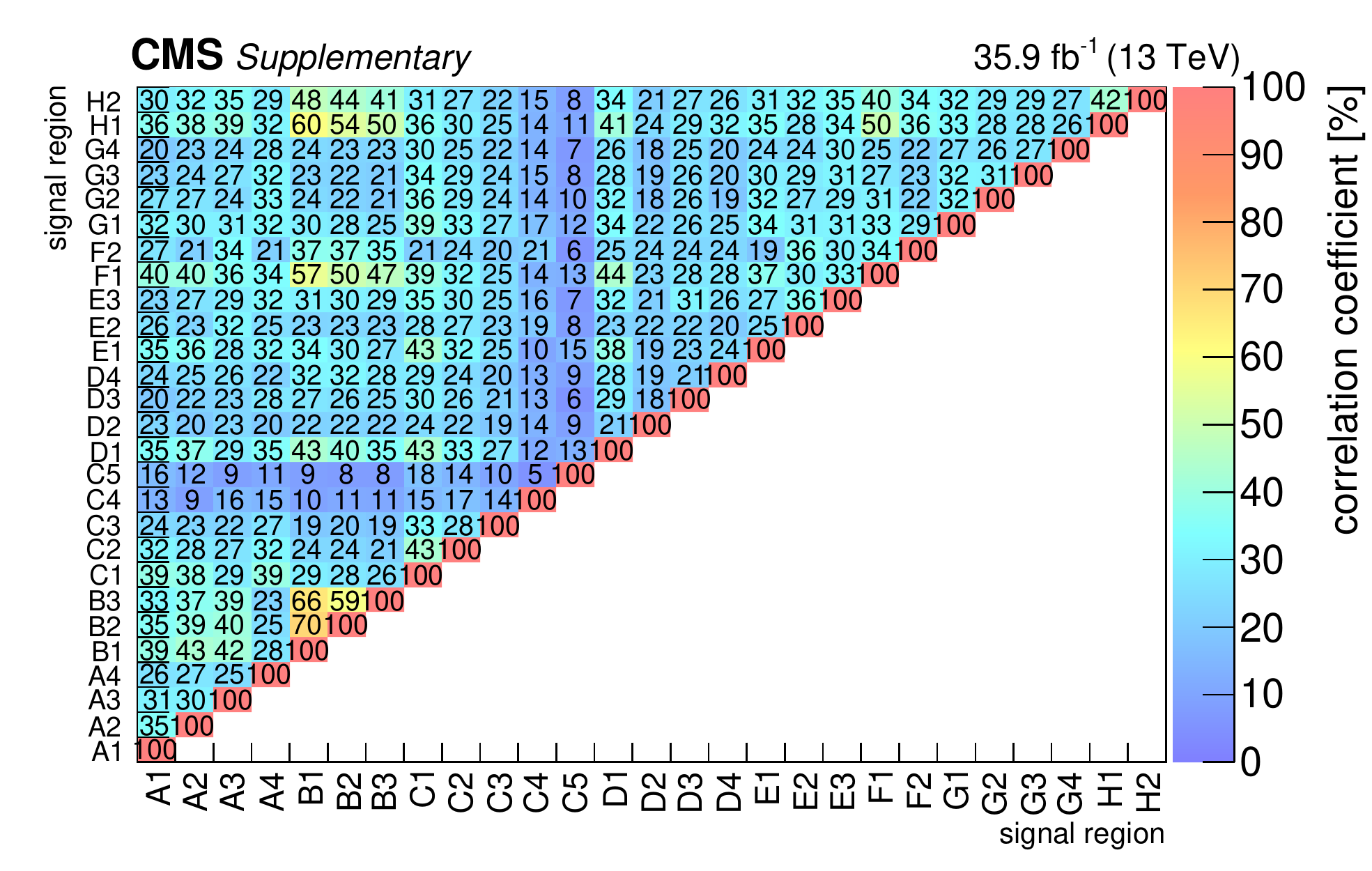}
\caption{\label{fig:corr1}Correlation matrix for the background predictions for the signal regions for the standard selection (in percent).  The labelling of the regions follows the convention of Fig.~4.}
\end{figure}

\begin{figure}[htb]
\centering
\includegraphics[width=\textwidth]{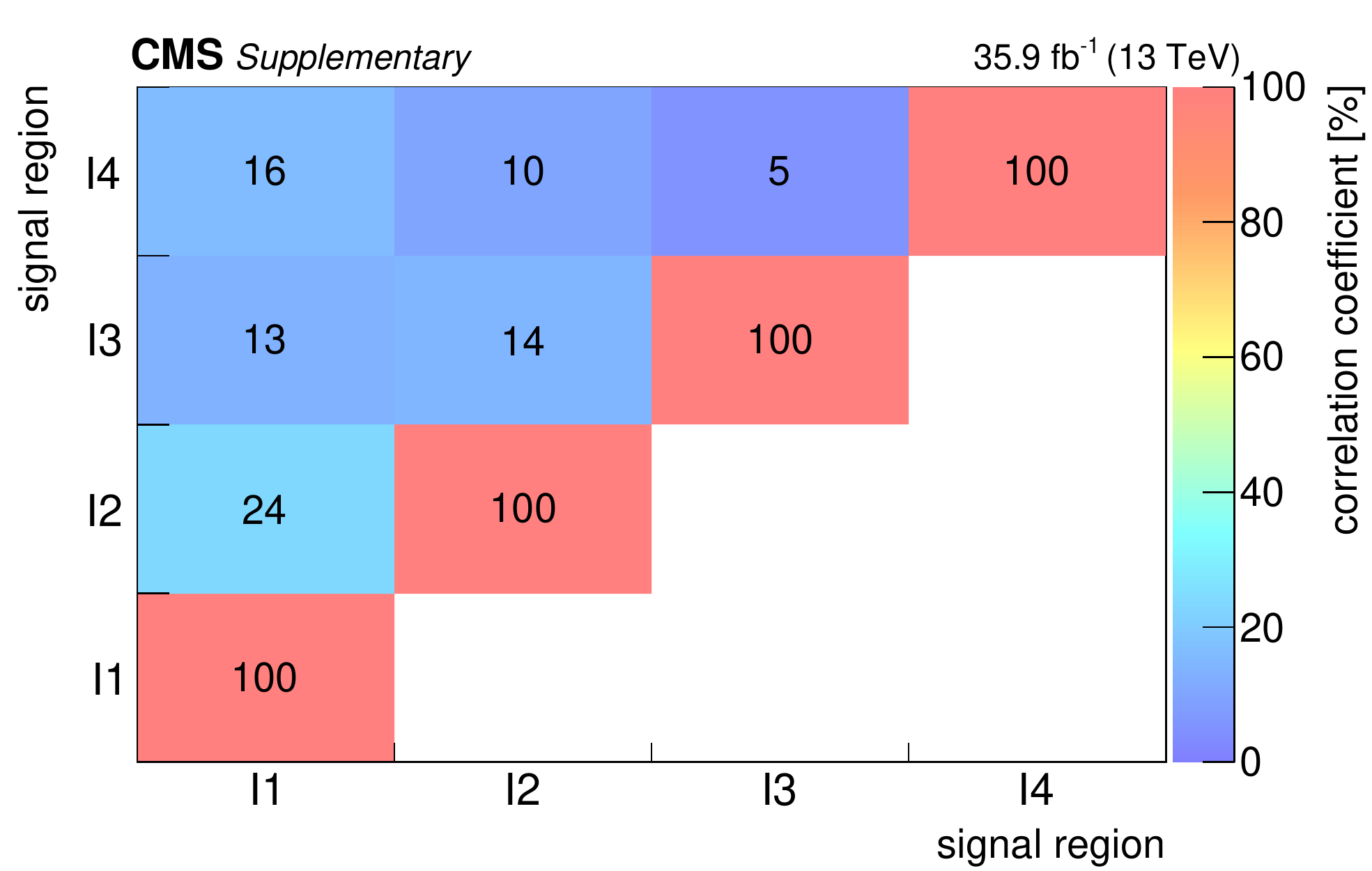}
\caption{\label{fig:corr2}Correlation matrix for the background predictions for the signal regions for the compressed selection (in percent).  The labelling of the regions follows the convention of Fig.~4.}
\end{figure}

The yields and background predictions of this search can be used
to confront scenarios for physics beyond the standard model (BSM) not considered in
this paper.  To facilitate such reinterpretations,
in Table~\ref{tab:aggregatedSR} we provide results for
a small number of inclusive aggregated signal regions.  The background
expectation, the event count, and the expected BSM yield in any one of
these regions can be used to constrain BSM hypotheses in a simple way.
In addition, we provide the correlation matrix for the background predictions in the full set of search regions (Figs.~\ref{fig:corr1} and~\ref{fig:corr2}).  This information can be used to exploit the full power of the analysis
by constructing a simplified
likelihood for a BSM model as described in Ref.~\cite{Collaboration:2242860}.

\begin{table}[htb]
\centering
\topcaption{\label{tab:aggregatedSR}Background predictions and yields in data corresponding to \Lint\ for aggregate signal regions.}
\resizebox{\textwidth}{!}{
\begin{tabular}{rrrrcccccc}
\hline
 \multirow{2}{*}{$N_\mathrm{J}$} & \multirow{2}{*}{$t_\text{mod}$} & $M_\mathrm{\ell b}$ & $E_\mathrm{T}^\text{miss}$ & Lost  & \multirow{2}{*}{1$\ell$ (top)} & 1$\ell$ (not & \multirow{2}{*}{$\Z\to\nu\bar{\nu}$} & Total & \multirow{2}{*}{Data} \\
  &  &  [\GeVns{}] &  [\GeVns{}] &  lepton &  &  top) &  & background &  \\
\hline
$\leq$3 &    $>$10 &            &    $>$600 & 0.6$\pm$0.5 & 0.3$\pm$0.3 & 1.7$\pm$0.5 & 0.8$\pm$0.5 & 3.4$\pm$0.9 & 4 \\
$\geq$4 & $\leq$0 & $\leq175$ &    $>$550 & 9.3$\pm$3.2 & 0.1$\pm$0.1 & 0.7$\pm$0.4 & 0.6$\pm$0.1 & 10.7$\pm$3.2 & 8 \\
$\geq$4 &    $>$10 & $\leq175$ &    $>$450 & 4.6$\pm$1.4 & 0.8$\pm$0.7 & 0.8$\pm$0.5 & 2.7$\pm$0.6 & 8.8$\pm$1.8 & 3 \\
$\geq$4 & $\leq$0 &     $>$175 &    $>$450 & 2.8$\pm$1.2 & 0.4$\pm$0.4 & 1.6$\pm$0.7 & 0.5$\pm$0.3 & 5.3$\pm$1.5 & 3 \\
$\geq$4 &     $>$0 &     $>$175 &    $>$450 & 0.3$\pm$0.2 & 0.1$\pm$0.1 & 0.7$\pm$0.3 & 0.7$\pm$0.2 & 1.9$\pm$0.5 & 2 \\
\hline
\multicolumn{3}{l|}{compressed region} &    $>$450 & 6.3$\pm$2.4 & 0.3$\pm$0.2 & 0.7$\pm$0.3 & 1.3$\pm$0.3 & 8.6$\pm$2.5 & 4 \\
\hline
\end{tabular}
}
\end{table}

\cleardoublepage \section{The CMS Collaboration \label{app:collab}}\begin{sloppypar}\hyphenpenalty=5000\widowpenalty=500\clubpenalty=5000\textbf{Yerevan Physics Institute,  Yerevan,  Armenia}\\*[0pt]
A.M.~Sirunyan, A.~Tumasyan
\vskip\cmsinstskip
\textbf{Institut f\"{u}r Hochenergiephysik,  Wien,  Austria}\\*[0pt]
W.~Adam, F.~Ambrogi, E.~Asilar, T.~Bergauer, J.~Brandstetter, E.~Brondolin, M.~Dragicevic, J.~Er\"{o}, M.~Flechl, M.~Friedl, R.~Fr\"{u}hwirth\cmsAuthorMark{1}, V.M.~Ghete, J.~Grossmann, J.~Hrubec, M.~Jeitler\cmsAuthorMark{1}, A.~K\"{o}nig, N.~Krammer, I.~Kr\"{a}tschmer, D.~Liko, T.~Madlener, I.~Mikulec, E.~Pree, D.~Rabady, N.~Rad, H.~Rohringer, J.~Schieck\cmsAuthorMark{1}, R.~Sch\"{o}fbeck, M.~Spanring, D.~Spitzbart, J.~Strauss, W.~Waltenberger, J.~Wittmann, C.-E.~Wulz\cmsAuthorMark{1}, M.~Zarucki
% ...
\vskip\cmsinstskip
% ...
1:~~Also at Vienna University of Technology, Vienna, Austria\\
% ...

\end{sloppypar}
\end{document}